%% file: main.tex
\documentclass[a4paper,11pt]{article}

\usepackage{jinstpub} 
\usepackage{lineno}

\usepackage{booktabs}

\usepackage{subfigure}
\usepackage{url}
\usepackage{hyperref}
\usepackage{siunitx}
\usepackage{adjustbox}
\usepackage{makecell}

\title{Beam-test evaluation of pre-production Low Gain Avalanche Detectors for the ATLAS High Granularity Timing Detector}

\input{HGTD_authors_sorted.tex}

\emailAdd{giulia.di.gregorio@cern.ch}
\emailAdd{stefano.manzoni@cern.ch}
\emailAdd{marion.missio@cern.ch}
\emailAdd{khuram.tariq@cern.ch}

\abstract{The High Granularity Timing Detector (HGTD) will be installed in the ATLAS experiment as part of the Phase-II upgrade for the High Luminosity-Large Hadron Collider (HL-LHC). It will mitigate pile-up effects in the forward region, and measure per bunch luminosity. The design of HGTD is based on Low Gain Avalanche Detector (LGAD) sensors. This paper presents the results of beam-test campaigns conducted at CERN and DESY in 2023 and 2024 on single LGADs from HGTD pre-production test structures, before and after neutron irradiation up to fluences of $2.5 \times 10^{15}~\mathrm{n_{eq}/cm^2}$. The tested LGADs can meet HGTD requirements in terms of charge collection, time resolution, and hit efficiency, even under HL-LHC end-of-life conditions, supporting their deployment in the final detector.}


\begin{document}
\maketitle
\flushbottom

\section{Introduction}
\input{Introduction.tex}

\section{Sensor} 
\input{Sensors.tex}

\section{Test beam set-up}
\input{Set-up.tex}

\section{Data analysis}
\input{Data_analysis}

\section{Sensor performance results}
\input{Sensor_results}

\section{Conclusions and outlook}
\input{Conclusions.tex}

\acknowledgments
\input{acknowledgments.tex}

\bibliography{ANA-HGTD-2025-01-INT1.bib}
\end{document}

%% file: HGTD_authors_sorted.tex
\author[u]{A.~Aboulhorma,}
\author[u]{M.~Ait~Tamlihat,}
\author[i,h]{H.M.~Alfanda,}
\author[n]{O.~Atanova,}
\author[n]{N.~Atanov,}
\author[q]{I.~Azzouzi,}
\author[b]{J.~Barreiro~Guimar\~{a}es~da~Costa,}
\author[z]{T.~Beau,}
\author[r]{D.~Benchekroun,}
\author[r]{F.~Bendebba,}
% ...

% ...
\affiliation[b]{Institute of High Energy Physics, Chinese Academy of Sciences, Beijing, China}
% ...
\affiliation[h]{State Key Laboratory of Dark Matter Physics, School of Physics and Astronomy, Shanghai Jiao Tong University, Key Laboratory for Particle Astrophysics and Cosmology (MOE), SKLPPC, Shanghai, China}
\affiliation[i]{State Key Laboratory of Dark Matter Physics, Tsung-Dao Lee Institute, Shanghai Jiao Tong University, Shanghai, China}

% ...
\affiliation[n]{affiliationiated with an international laboratory covered by a cooperation agreement with CERN}

% ...
\affiliation[q]{Moroccan Foundation for Advanced Science Innovation and Research (MAScIR), Rabat, Morocco}

\affiliation[r]{Facult\'e des Sciences Ain Chock, Universit\'e Hassan II de Casablanca, Casablanca, Morocco}
% ...
\affiliation[u]{Facult\'e des sciences, Universit\'e Mohammed V, Rabat, Morocco}
% ...
\affiliation[z]{LPNHE, Sorbonne Universit\'e, Universit\'e Paris Cit\'e, CNRS/IN2P3, Paris, France}

% ...

%% file: Introduction.tex
\label{sec:intro}

The High Granularity Timing Detector (HGTD)~\cite{2091129} will be installed in the ATLAS experiment~\cite{ATLAS:2008xda} during the so-called Phase-II upgrade for the High-Luminosity period of the LHC (HL-LHC)~\cite{Apollinari:2017lan}. The HGTD pixel detector will be introduced in the ATLAS detector at the gap region between the tracker and the end-cap calorimeters at a distance of $\pm$3.5 m from the interaction point covering the pseudorapidity range of 2.4 $< |\eta| < $ 4.0. The main goal of this detector is to mitigate the pile-up issue due to the increase in the instantaneous luminosity during the high-luminosity phase of the LHC by combining the time measurements of the hits with the spatial information provided by the Inner Tracker. In addition, the HGTD detector offers unique capabilities for online and offline luminosity determination, an important requirement for precision physics measurements. 

The HGTD will consist of 3.16 million silicon-based Low Gain Avalanche Detectors (LGAD)~\cite{Pellegrini:2014lki}. LGADs are n-in-p diodes structures with an additional p+ type gain layer, which enhances the rise time and the signal-to-noise ratio, allowing therefore for excellent time resolution. LGAD sensors have been extensively studied, before and after irradiation, in the past years~\cite{Allaire:2018bof,Agapopoulou:2020gnf,Lange:2017pxs,Agapopoulou:2022vxk}. Due to the expected high radiation levels, the HGTD is designed to withstand a 1~MeV neutron-equivalent fluence of up to $2.5\times10^{15}\text{n}_{\mathrm{eq}}/\text{cm}^{2}$ and a total ionizing dose of 2~MGy, estimated at a distance of 120 mm from the beam pipe~\cite{2091129}. The most exposed regions are expected to receive radiation damage beyond this level and will therefore require replacement.
In order to reach the HGTD design performance, the LGADs must meet targets regarding time resolution of \SI{40}{\pico\second} (\SI{50}{\pico\second}) per hit, collected charge $>$\SI{15}{\femto\coulomb} ($>$\SI{4}{\femto\coulomb}) and hit efficiencies $>$ 97\% ($>$ 95\%) in the central part of the pad at the start (end) of their lifetime. These requirements should be achieved taking into account a discriminator threshold of about \SI{2}{\femto\coulomb} of the ALTIROC, front-end ASIC foreseen for HGTD~\cite{deLaTaille:2018jyb}.
%

In beam tests~\cite{Agapopoulou:2022vxk,Ali:2023roa} as well as in laboratory measurements~\cite{Garcia:2021ujp}, it was established that the addition of carbon in the gain layer helps reducing the acceptor removal rate with irradiation, leading to improved radiation hardness and a lower operating voltage needed to collect the same charge of the sensors~\cite{Ferrero:2018fen,Padilla:2020sau}. This is extremely important at high fluences, where LGADs need a high bias voltage to maintain their performance. The maximal bias voltage is limited by the destructive discharge events in the sensor, which limit the safe average electric field in the sensor to values below 11 V/$\mu$m~\cite{Beresford:2023egc}. For a 50-$\mu$m-thick HGTD LGAD sensor, this corresponds to a safe operating bias voltage of about 550 V.
%

The present paper describes the performance of pre-production LGAD sensors with carbon enriched gain layer, to qualify the sensor design for use in the HGTD detector. These pre-production sensors are advanced prototypes that are close to their final design and manufactured in limited quantities for validation before mass production. The tested LGADs are described in Section~\ref{sec:sensors}. The response of the LGADs to particles has been tested in test beam facilities with an experimental set-up presented in Section~\ref{sec:set-up}. The collected data are analysed according to the method described in Section~\ref{sec:data_analysis}. The results of these tests are presented in Section~\ref{sec:results}.

%% file: Sensors.tex
\begin{figure}[htbp]
\centering
\subfigure[USTC--IME]{
    \includegraphics[width=0.45\textwidth]{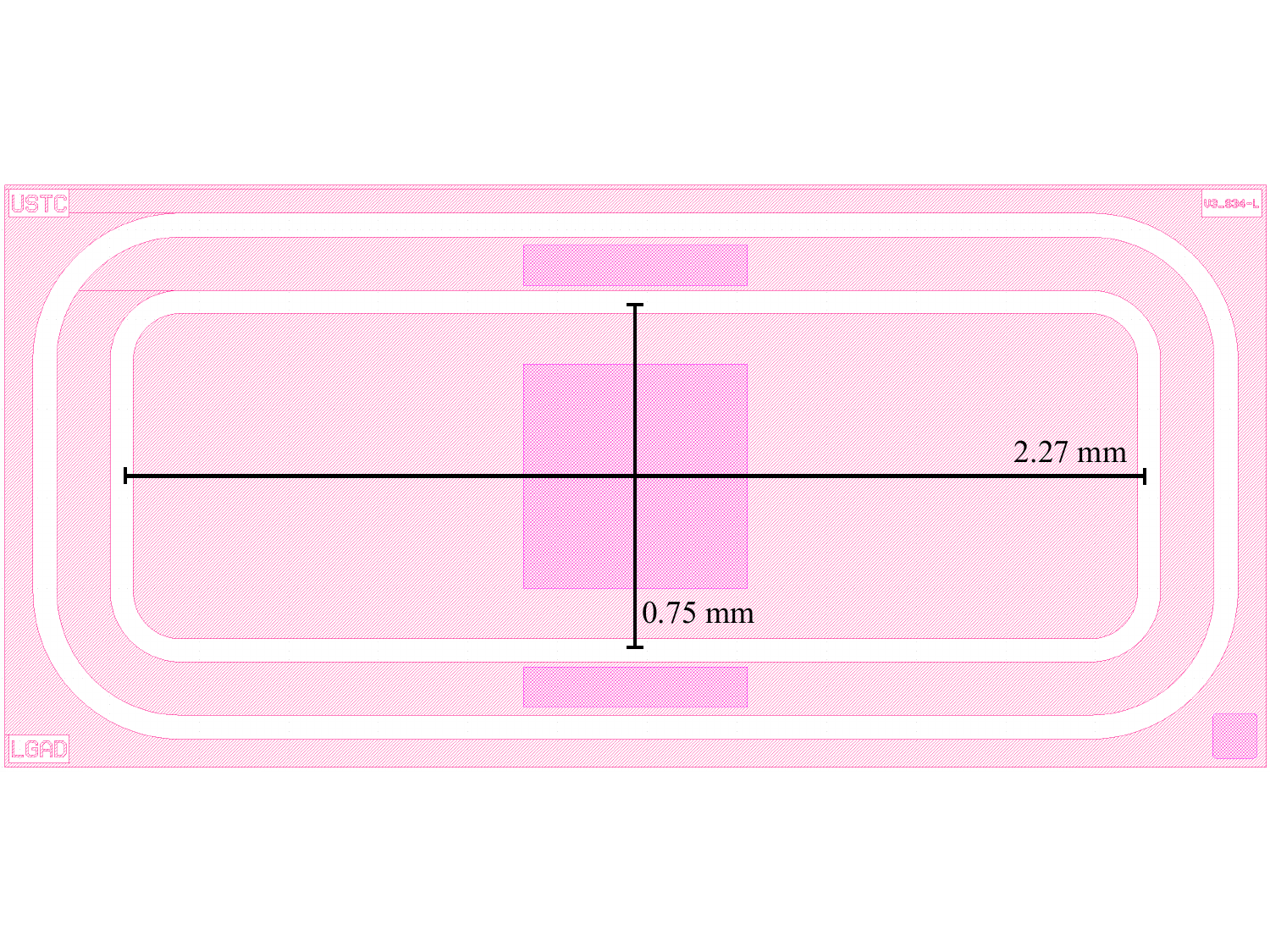}}
\qquad
\subfigure[IHEP--IME]{
    \includegraphics[width=0.45\textwidth]{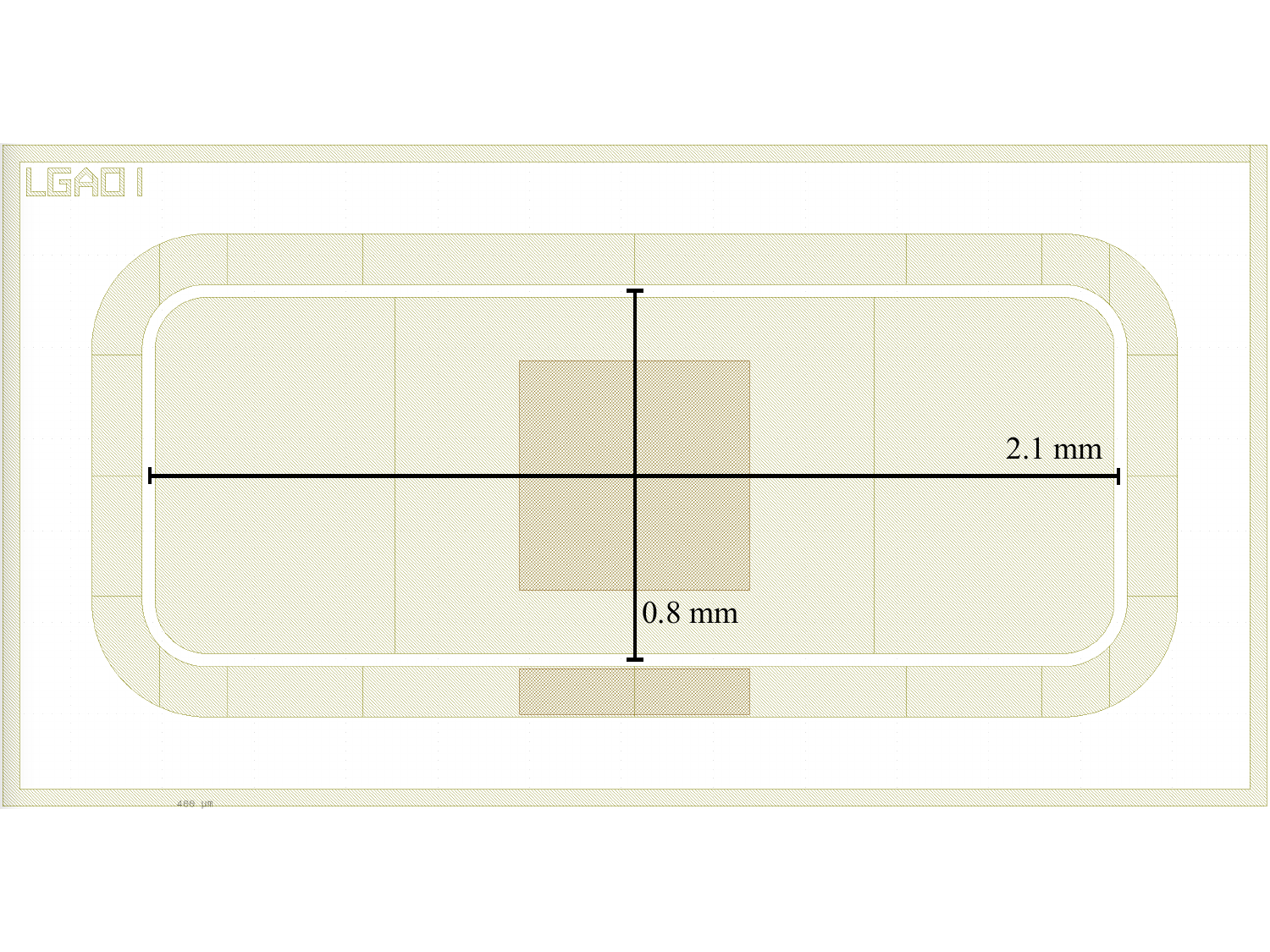}}
\caption{Layout of the single-pad LGAD of the QC-TS from the USTC-IME and IHEP-IME designs.}
\label{fig:QC-TS_layout}
\end{figure}

\label{sec:sensors}
The HGTD sensors are manufactured by the Institute of Microelectronics (IME)~\cite{IME:CAS_overview} of the Chinese Academy of Science in China, based on two different designs, one by the Institute of High Energy Physics (IHEP)~\cite{Li:2021ouc} and the other by the University of Science and Technology of China (USTC)~\cite{Li:2022ukj}. The LGAD sensors are produced on 8-inch wafers. The total thickness of the sensors is 775$\pm$5 $\mu$m, while the active thickness is 50$\pm$5 $\mu$m. Each sensor is a 15 $\times$ 15 array of 1.3 $\times$ 1.3 mm$^2$ LGAD pads, for an active area of 380.25 mm$^2$. The sensors have a boron implant gain layer with an additional co-implantation of carbon, which improves the radiation hardness through smaller initial acceptor removal. \\
To check the performance of the sensors that will be installed in the detector, an equal amount of Quality Control-Test Structures (QC-TS) is produced along the sensors. In this way, the quality and uniformity of the wafers can be monitored and wafer-level technology parameters can be extracted during the production, allowing the HGTD Collaboration to study wafer and sensor properties in detail, without interfering with the hybridisation and module-production work flow. Using QC-TS, bad wafers or issues during the production can be spotted by monitoring testing parameters over large periods of time. The QC-TS consists of several structures, including a single-pad LGAD, as show in Fig.~\ref{fig:QC-TS_layout} for IHEP and USTC, respectively. The sensor has an elongated dimension of 2.27 $\times$ 0.75 mm$^2$ and 2.1 $\times$ 0.8 mm$^2$ for USTC and IHEP sensors, respectively. This dimension corresponds to the same area of the LGAD pixels in the sensors that will be mounted in the detector. Differently, the dimension of the gain layer is 2.222 $\times$ 0.704 mm$^2$ and 2.062 $\times$ 0.750 mm$^2$. The study of the LGADs, presented in this paper, is based on the single-pad LGADs from the QC-TS tested in several test-beam campaigns in 2023 and 2024 at CERN and DESY.

The tested sensors are from the so-called "pre-production", a small fraction of sensors with the final design, produced by the chosen vendor and tested before launching the full production. More specifically, the pre-production sensors are categorized as either early pre-production or core pre-production. Both early pre-production and core pre-production sensor have a reduced physical thickness of 300$\pm$5 $\mu$m\footnote{The sensor substrate thickness was changed between the pre-production and the final production to increase the resistance of the sensor and ASIC to thermal cycles. The sensor substrate thickness does not affect the performance since the active thickness remains unchanged.}. Except for the thickness, the core pre-production sensors have identical design, specifications, active thickness and requirements as the final production sensors. In contrast, the early pre-production sensors differ in several parameters. In particular, the early pre-production sensors studied in this paper do not have the under bump metallisation (UBM) and, only the IHEP designed sensors have an increased thickness of 775$\pm$5 $\mu$m. Both, the pad metallisation and the physical thickness are not expected to significantly affect the sensor performance.

To study the LGAD performance after irradiation, the sensors were exposed to fluences up to $2.5\times10^{15}~\text{n}_{\mathrm{eq}}/\text{cm}^{2}$ at the TRIGA reactor~\cite{Snoj:2012dib,Ambrozic:2017} in Ljubljana, Slovenia with fast neutrons. Table~\ref{tab:sensors_list} lists the LGAD sensors measured in the test beams, including the vendor, physical dimensions, and the irradiation fluence.  In the table the first column includes the device name assigned to each sensor for easier reference in the text: the sensor name is a concatenation of the sensor design (IHEP or USTC), the wafer number and the fluence. For instance, USTC-W7-2.5 refers to a USTC LGAD from wafer 7, exposed to a fluence of $2.5\times10^{15}~\text{n}_{\mathrm{eq}}/\text{cm}^{2}$. In one case, to distinguish two sensors, the position of the sensor in the wafer has been added in parenthesis.  \\

\begin{table}[htbp]
\centering
\caption{List of IHEP-IME and USTC-IME LGAD sensors studied in the 2023 and 2024 beam-test campaigns, including irradiation fluence, type, physical dimensions, and test facility.}
\smallskip

\label{tab:sensors_list}
\footnotesize
\begin{tabular}{l|c|c|c|c|c|c}
\toprule
\textbf{Device name} & \textbf{Design} & \textbf{UBM} &
\textbf{Thickness [$\mu$m]} & \textbf{Dimension [mm$^2$]} &
\textbf{Fluence[$\mathrm{n_{eq}/cm^2}$]} & \textbf{Tested at} \\
\midrule
USTC-W2-0     		& USTC &  no & 300$\pm$5  & 2.27$\times$0.75 & 0 & CERN \\
USTC-W2-0.8   		& USTC &  no & 300$\pm$5  & 2.27$\times$0.75 & 0.8$\times10^{15}$ & CERN \\
USTC-W2-1.5   		& USTC &  no & 300$\pm$5  & 2.27$\times$0.75 & 1.5$\times10^{15}$ & CERN \\
USTC-W2-2.5   		& USTC &  no & 300$\pm$5  & 2.27$\times$0.75 & 2.5$\times10^{15}$ & CERN \\
USTC-W24-0.8  		& USTC &  no & 300$\pm$5  & 2.27$\times$0.75 & 0.8$\times10^{15}$ & CERN \\
USTC-W24-1.5  		& USTC &  no & 300$\pm$5  & 2.27$\times$0.75 & 1.5$\times10^{15}$ & CERN \\
USTC-W24-2.5  		& USTC &  no & 300$\pm$5  & 2.27$\times$0.75 & 2.5$\times10^{15}$ & CERN \\
USTC-W7-0      		& USTC &  yes & 300$\pm$5  & 2.27$\times$0.75 & 0 & CERN \\
USTC-W15(P49)-2.5 	& USTC &  yes & 300$\pm$5  & 2.27$\times$0.75 & 2.5$\times10^{15}$ & CERN \\
USTC-W15(P1)-2.5 	& USTC &  yes & 300$\pm$5  & 2.27$\times$0.75 & 2.5$\times10^{15}$ & CERN \\
\midrule
IHEP-W2-1.5 		& IHEP &  no & 775$\pm$5   & 2.1$\times$0.8 & 1.5$\times10^{15}$ & DESY \\
IHEP-W2-2.5 		& IHEP &  no & 775$\pm$5   & 2.1$\times$0.8 & 2.5$\times10^{15}$ & DESY \\
IHEP-W10-0 			& IHEP &  yes & 300$\pm$5  & 2.1$\times$0.8 & 0 & CERN \\
IHEP-W16-0 			& IHEP &  yes & 300$\pm$5  & 2.1$\times$0.8 & 0 & CERN \\
IHEP-W16-1.5 		& IHEP &  yes & 300$\pm$5  & 2.1$\times$0.8 & 1.5$\times10^{15}$ & CERN \\
IHEP-W16-2.5 		& IHEP &  yes & 300$\pm$5  & 2.1$\times$0.8 & 2.5$\times10^{15}$ & CERN \\
IHEP-W10-2.5 		& IHEP &  yes & 300$\pm$5  & 2.1$\times$0.8 & 2.5$\times10^{15}$ & CERN \\ 
\bottomrule
\end{tabular}
\end{table}

\begin{figure}[htbp]
\centering

\subfigure[IHEP unirradiated]{
    \includegraphics[width=0.45\textwidth]{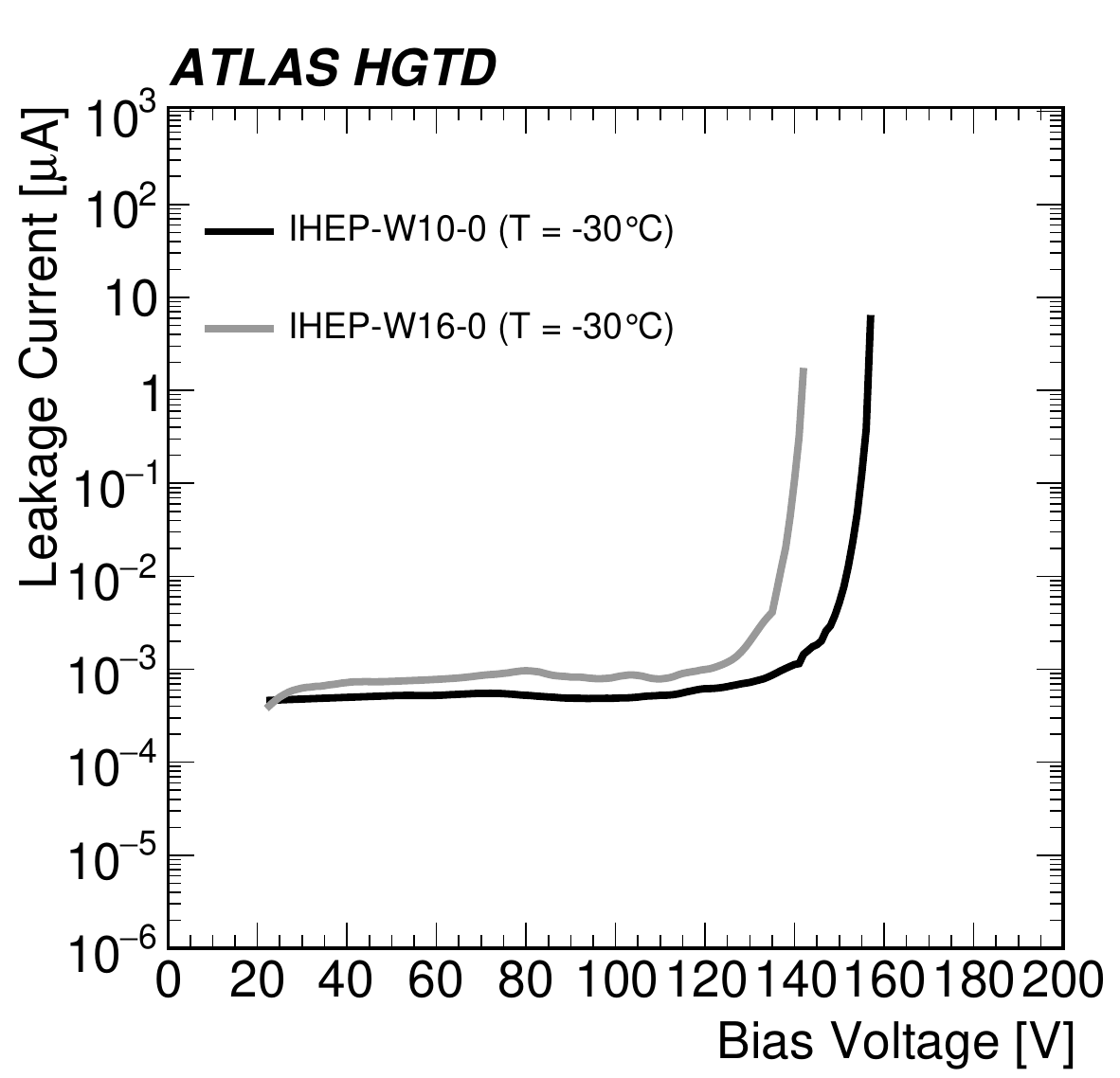}}
\qquad
\subfigure[IHEP irradiated]{
    \includegraphics[width=0.45\textwidth]{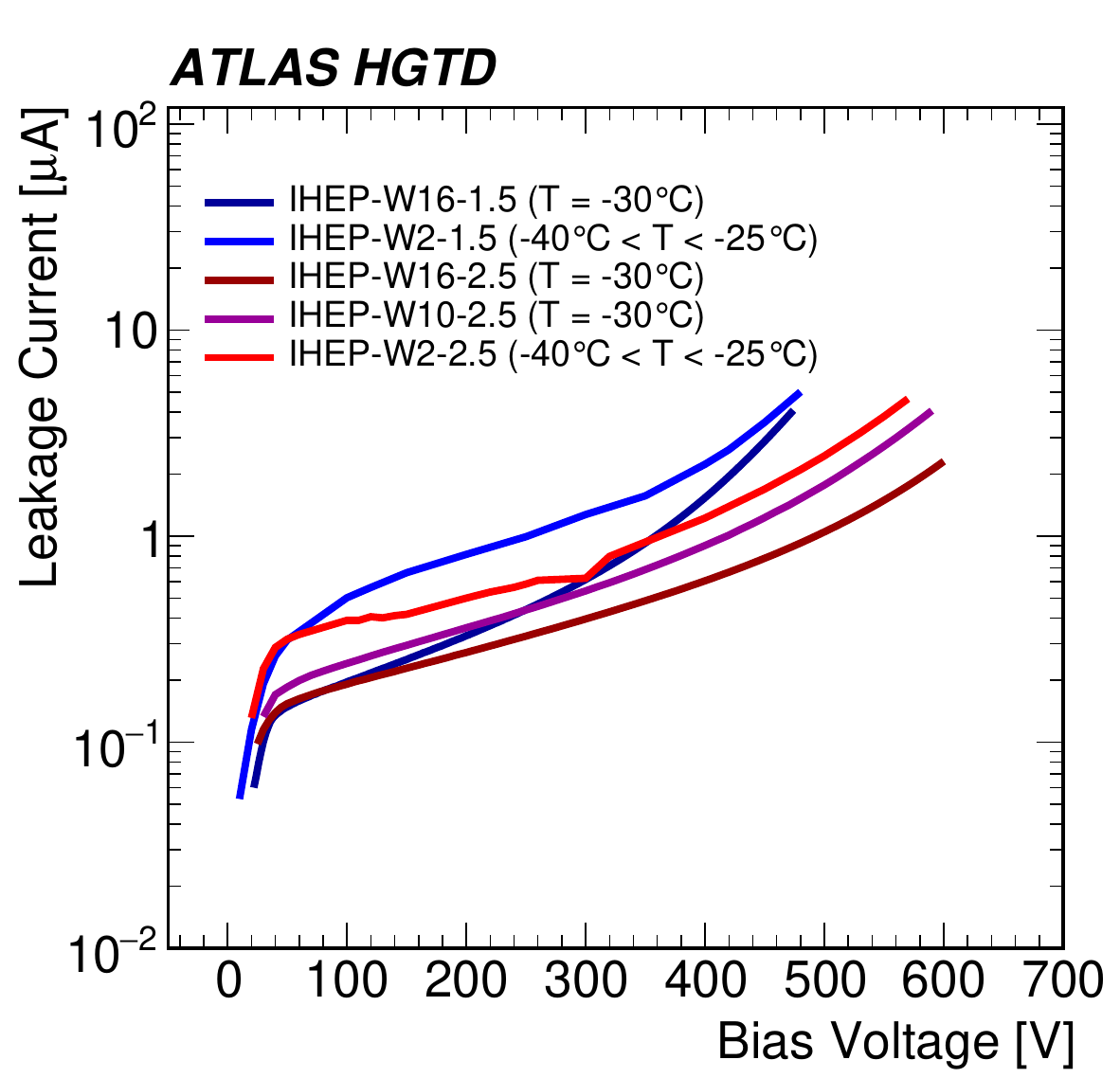}}

\subfigure[USTC unirradiated]{
    \includegraphics[width=0.45\textwidth]{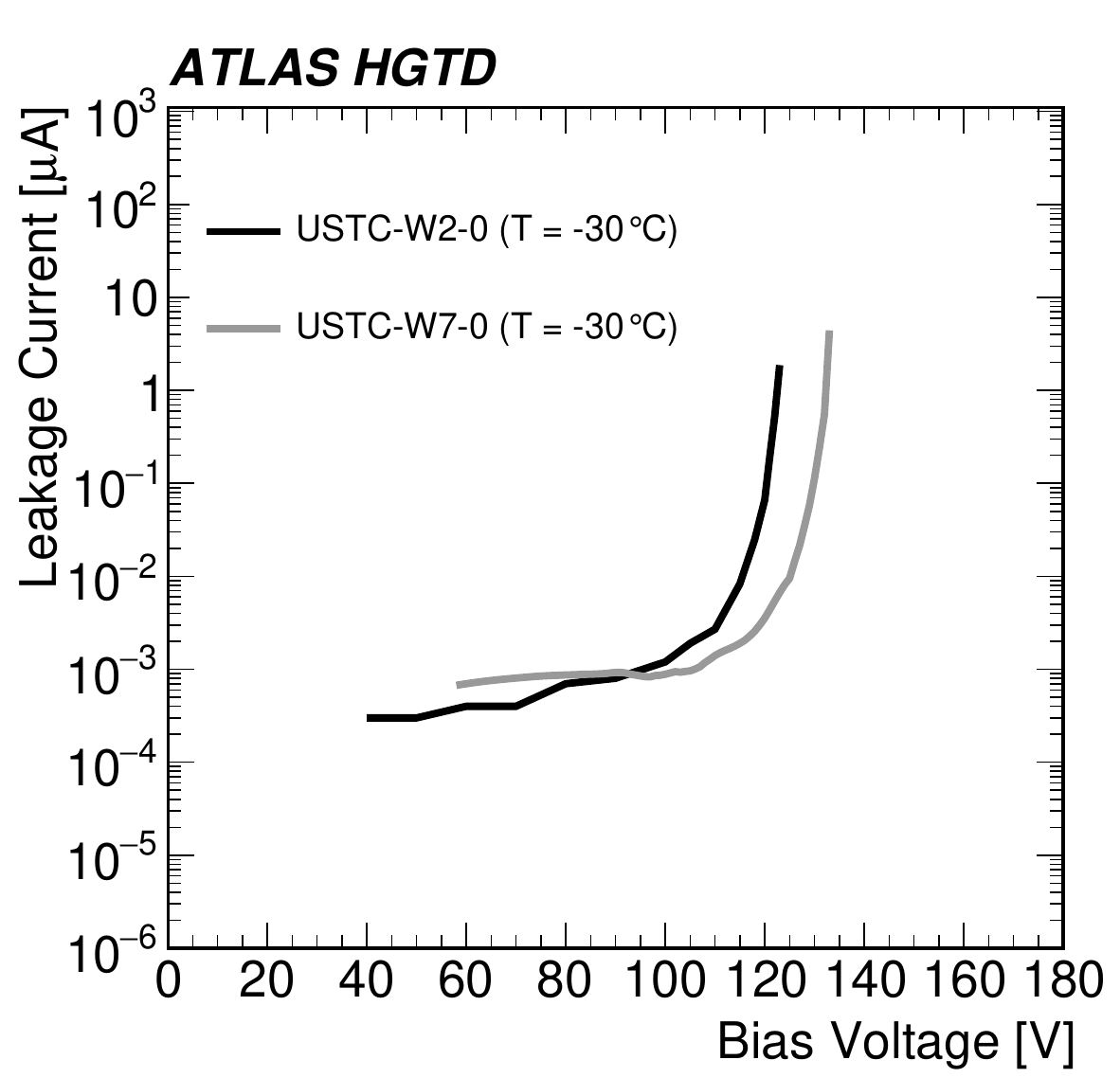}}
\qquad
\subfigure[USTC irradiated]{
    \includegraphics[width=0.45\textwidth]{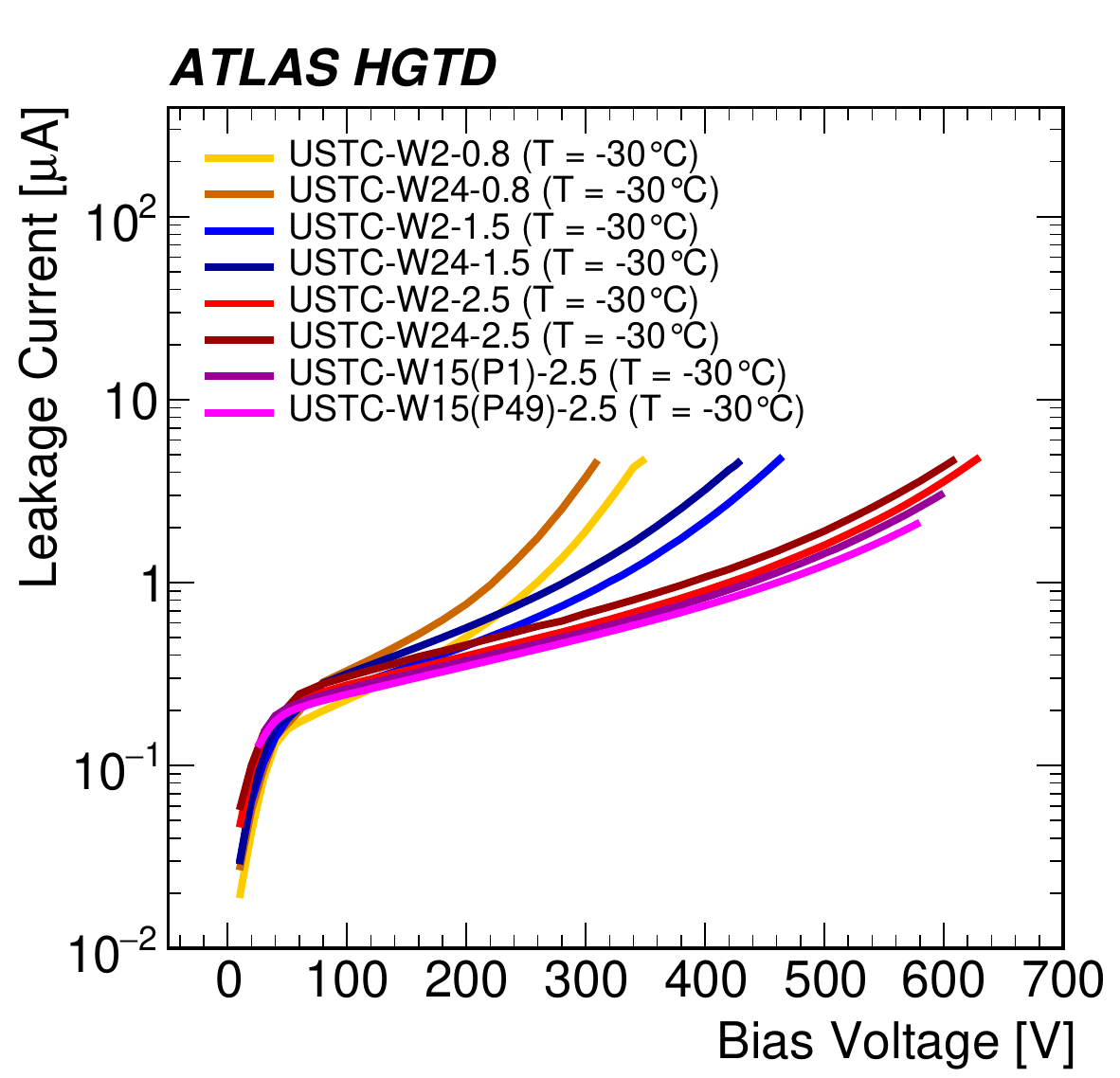}}

\caption{Leakage current as a function of the bias voltage measured at $-30^\circ$C for all tested sensors.}
\label{fig:IV_curves}
\end{figure}

Figure \ref{fig:IV_curves} shows the leakage current-voltage (I-V) characteristics at around -30 $^{\circ}$C for the tested sensors. At the same irradiation fluence, the sensors exhibit similar performance.

%% file: Set-up.tex
\begin{figure}[htbp]
\centering
\subfigure[]{
    \includegraphics[width=0.45\textwidth]{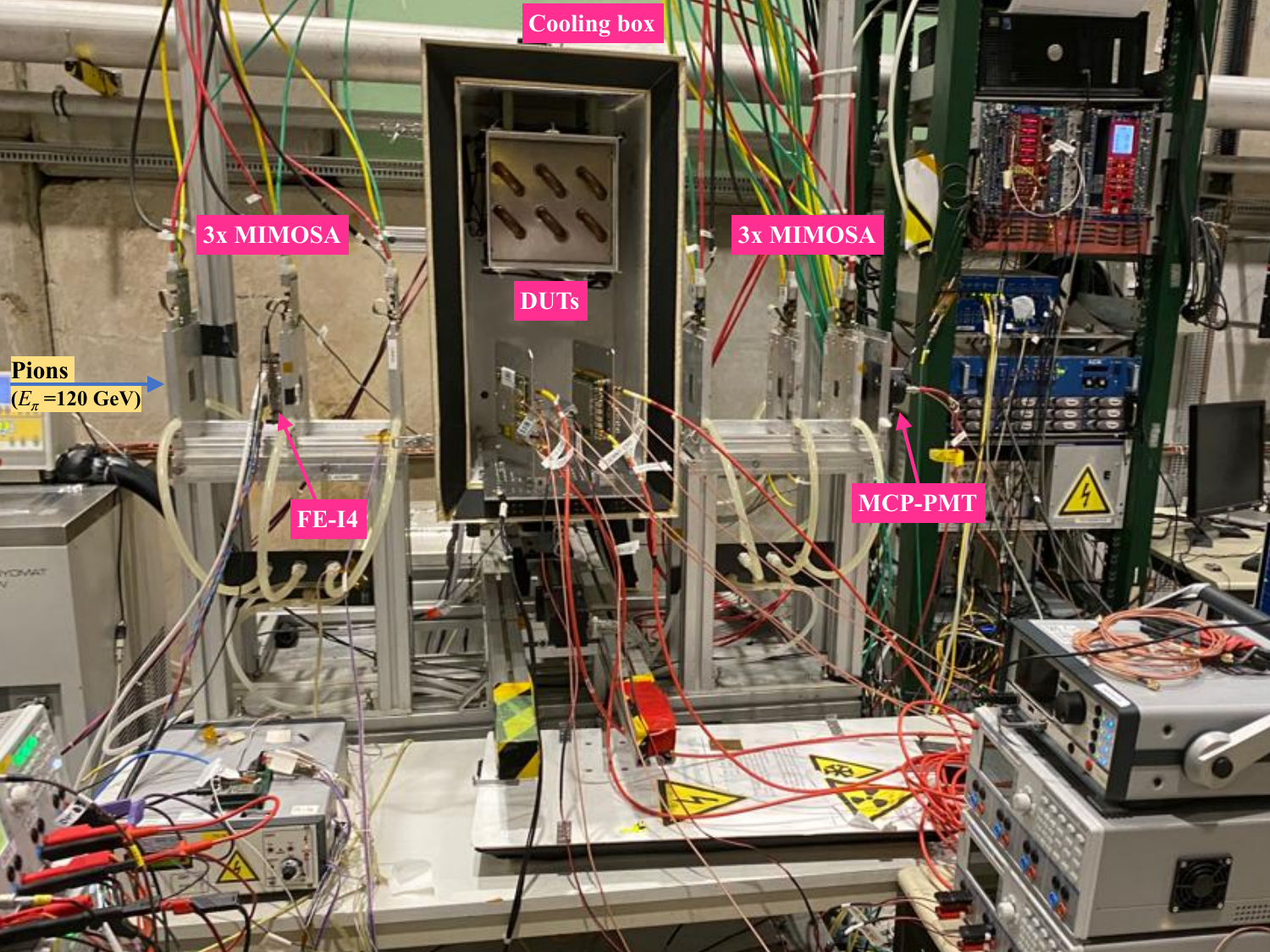}}
\qquad
\subfigure[]{
    \includegraphics[width=0.45\textwidth]{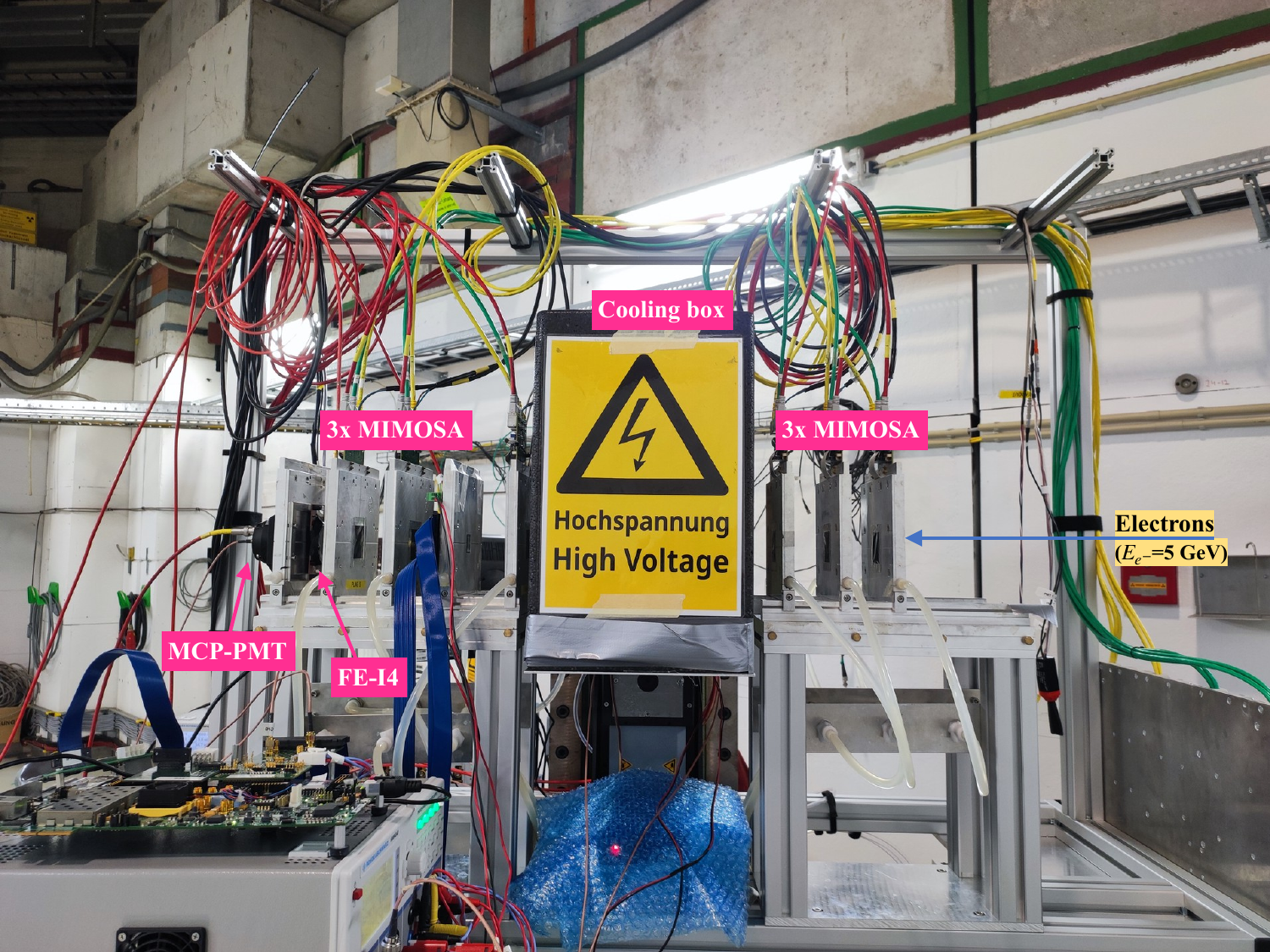}}
\caption{Pictures of the test beam set-up at CERN(a) and at DESY(b).}
\label{fig:TB_setup}
\end{figure}

\label{sec:set-up}
The measurement of the HGTD pre-production LGADs were conducted during two beam-test campaigns at CERN SPS~\cite{SPS_H6} H6A beam line using a high-momentum \SI{120}{\giga\electronvolt} pion beam and one campaign at the DESY II test beam facility~\cite{Diener:2018qap} using a \SI{5}{\giga\electronvolt} electron beam. The data-taking set-ups at CERN and DESY were similar, with the main difference being the cooling systems used for the devices under test (DUTs). At CERN, a chiller was employed, ensuring stable operation at -30$^{\circ}$C, whereas at DESY cooling was provided by dry ice packs, and measurements were performed in the range of -40$^{\circ}$C to -25$^{\circ}$C. 
Figure \ref{fig:TB_setup} shows the set-up at CERN and at DESY. 

\begin{figure}[htbp]
	\centering
	\includegraphics[width=0.4\textwidth]{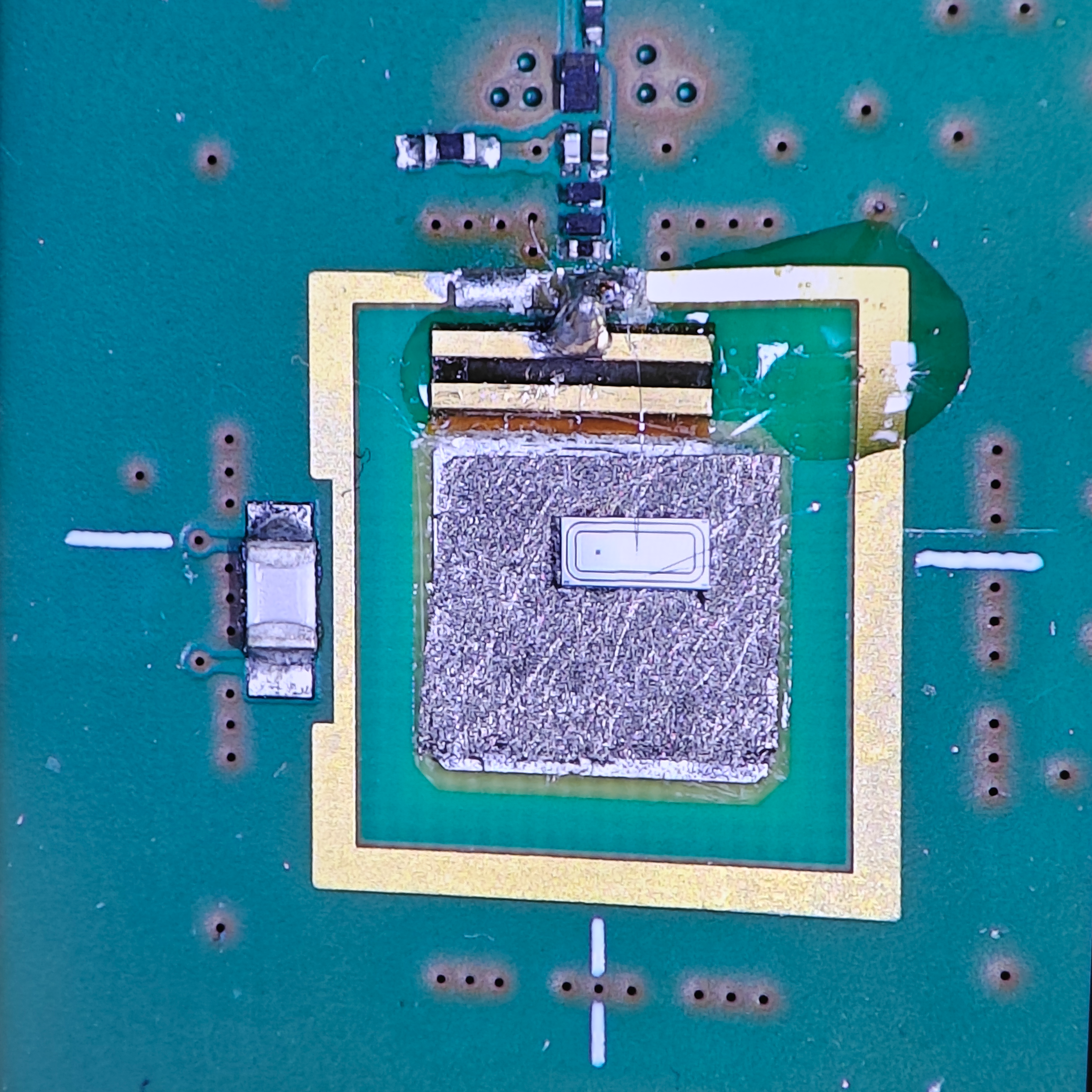}
	\caption{Picture of the QC-TS single pad LGAD mounted on a readout board.}
	\label{fig:QC-TS_readout}
\end{figure}

The LGADs were mounted on custom readout boards developed at the University of Santa Cruz ~\cite{Agapopoulou:2022vxk,Galloway:2017gfx} with an on-board amplification stage, as shown in Figure \ref{fig:QC-TS_readout}. The first internal amplifier was followed by an external commercial second-stage 20 dB amplifier to give a trans-impedance of 4700 $\Omega$. The second stage amplifier was placed either inside the cooling box or outside at room temperature. To account for variations in results due to changes in the amplifier trans-impedance with temperature, a 10\% systematic uncertainty was assigned to the collected charge and time reconstruction. This uncertainty is based on a comparison of benchmark measurements with the second-stage amplifier inside versus outside the cooling box. 

The time resolution of each DUT was estimated using a Micro-Channel Plate Photomultiplier Tube (MCP-PMT) HPK R3809U-50~\cite{Hamamatsu:R3809U-50} as a time reference. The MCP-PMT was calibrated using coincidences between the MCP-PMT and two DUTs. The time resolution was extracted by considering only the overlapping areas, yielding a measured value of 10.6 $\pm$ 2.2 ps at an operating voltage of 2650 V. The results were cross-checked in dedicated runs at SPS using two additional MCP-PMTs in coincidence, which gave a measured resolution of about 13 ps, in agreement with the previous measurement within uncertainties. \\

\begin{figure}[htbp]
	\centering
	\includegraphics[width=0.9\textwidth]{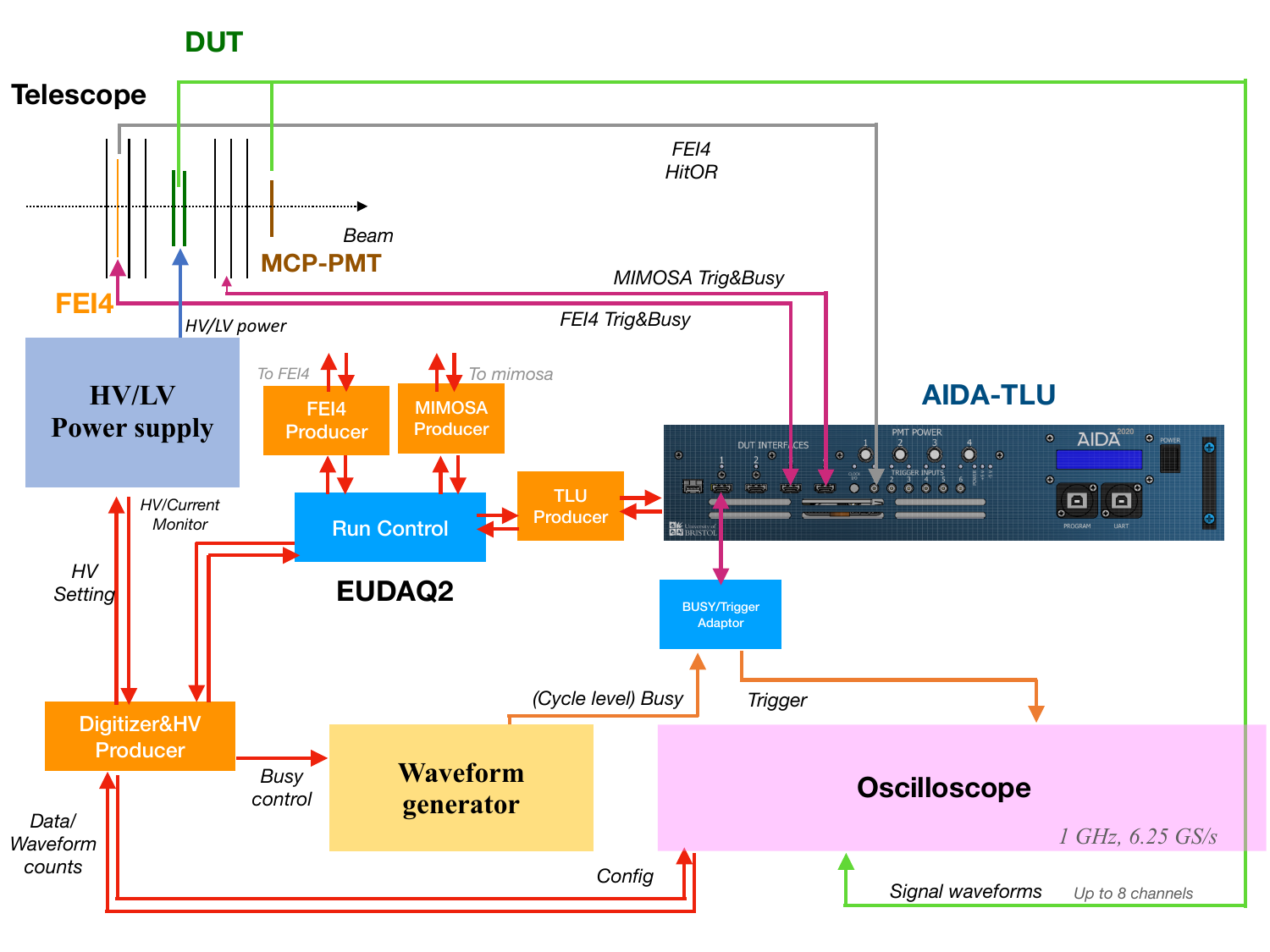}
	\caption{Simplified schematic of the DAQ system used during the test-beam campaigns at CERN and DESY. The TLU receives triggers from the FEI4 HitOr signal (grey line). All DUTs are interfaced to the TLU via LVDS signals (magenta lines). The oscilloscope is connected through an adapter board that converts LVDS to TTL and splits the trigger and busy signals (orange lines). The green lines represent the sensor and MCP signals connected to the oscilloscope. Red lines indicate all connections routed through the internal network. Finally, the blue line shows the connections between the high-voltage (HV) and low-voltage (LV) power supplies and the sensor boards.}
	\label{fig:TB_DAQ}
\end{figure}

An eight-channel oscilloscope was used to sample the waveforms from both the DUTs and the MCP-PMT, as shown in Figure \ref{fig:TB_DAQ}. The oscilloscope has a 1 GHz bandwidth, a 6.25 GS/s sample rate with 8-bit resolution and a memory depth allowing the acquisition of up to 50k samples per channel per trigger cycle. 
The setup also included a EUDET-type telescope based on six MIMOSA~\cite{Jansen:2016bkd} pixel planes (three placed on each side of the DUT) to reconstruct the trajectory of the incident charged particles. The telescope provided the hit position of each track in the DUT frame, enabling efficiency and timing studies.\\
Triggering was provided by a FE-I4~\cite{Benoit_2016} readout chip–based module, which initiated the acquisition of both oscilloscope waveforms and telescope data. The FE-I4 was configured to accept signals only in a region of interest (ROI) matching the DUT geometries. Trigger logic and distribution were managed by a programmable Trigger Logic Unit (TLU)~\cite{Baesso:2019smg}. Busy signals were propagated within the system: in particular, a waveform generator was used to rise a busy signal during the oscilloscope readout, while the TLU also distributed busy states from the telescope and FE-I4. In this way, triggers were vetoed during dead times, ensuring synchronized and loss-free data collection. In addition, the FE-I4 signal was recorded and subsequently used in the offline analysis.

%
The entire setup was integrated under a DAQ framework based on the EUDAQ2 software~\cite{Liu:2686241}, which ensured synchronization of the TLU, telescope, and FE-I4 oscilloscope  data streams. In addition, the HV power supply of the DUTs was controlled through EUDAQ2, enabling unsupervised and automatic data-taking over different voltage configurations. Figure~\ref{fig:TB_DAQ} summarises the DAQ system workflow implemented in the test-beam campaigns.

During the June 2024 test beam campaign at the SPS, measurements were performed at different angles. A rotation stage with stepping motor was installed inside the cooling box to remotely control the rotation of the sensor with respect to the incident beam. 
In addition to the standard measurement with the beam perpendicular to the DUT (0$^{\circ}$), data were also collected at 6$^{\circ}$, 12$^{\circ}$, and 18$^{\circ}$. Considering the HGTD geometry and the position inside the ATLAS detector, the maximum angle at which the particles can hit the detector is around 10$^{\circ}$.


%% file: Data_analysis.tex
\label{sec:data_analysis}
Two independent systems were used to collect the data analyzed in this study: the EUDET-type telescope and FE-I4 plane, which provided track information, and the oscilloscope, which recorded the waveforms from the LGAD and MCP-PMT detectors. This section describes the general methodology used to reconstruct and process the information from these two chains, and how it is used to derive the physical quantities for the analysis.

\subsection{Digitizer Data Processing}
\label{sec:digitizer_data_processing}

The waveform processing procedure followed a multi-step approach. First, the binary data from the oscilloscope were converted into a ROOT-format ntuple containing raw waveform samples for each channel. Each waveform was then analyzed to determine the signal polarity, maximum and minimum amplitudes, and the corresponding start and stop points of the signal. These parameters were used to verify that the entire pulse was contained within the acquisition window.

Next, the LGAD signal baseline (pedestal) and noise level were estimated using a region preceding the pulse onset (typically between 10\% and 90\% of the pre-pulse window). The pedestal was computed as the mean of this pre-pulse region, and the noise was determined as the standard deviation obtained from a Gaussian fit to the distribution of sample amplitudes within this region. The pedestal value was subtracted from all waveform samples on an event-by-event basis, and signal characteristics were recalculated post-subtraction.

The MCP-PMT waveforms were processed following a similar procedure, with adjustments to account for their distinct signal characteristics. MCP-PMT pulses exhibit faster rise times and higher amplitudes compared to the LGAD signals, and the noise level is generally lower due to the high intrinsic gain of the MCP-PMT. The baseline and noise were computed using the same pre-pulse region definition, and pedestal subtraction was applied identically. However, the integration window used to determine the collected charge was shorter (typically 1–2 ns) to match the narrower pulse width of the MCP-PMT signals. The waveform properties were stored in the same ROOT-file ntuple using vector branches for the DUT (LGAD) and MCP-PMT signals, allowing for synchronized event-by-event analysis.

Following baseline correction, several waveform properties were extracted, including the signal amplitude, rise time, collected charge, signal-to-noise ratio (SNR), and Time of Arrival (TOA). The amplitude was defined as the global maximum of the waveform. A cut on the signal amplitude was used to remove events with low amplitude, which helps reduce noise contributions. The cut value is 40 mV and 10 mV for unirradiated and irradiated sensors, respectively. The charge, $q$, was determined by dividing the integral of the pulse by the trans-impedance of the read-out board, $R_{\text{b}}=4700 \Omega$, and the gain of the voltage amplifier, $G_{\text{ampl}} = 10^{12}$: 
\begin{equation}
q = \frac{\int_{\text{start}}^{\text{stop}} A \, dt}{R_{\text{b}} \times G_{\text{ampl}}}\ .
\end{equation}
The TOA was extracted using the Constant Fraction Discriminator (CFD) method, defined as the point where the signal crosses a fixed fraction (fCFD) of its peak amplitude. The TOA value was chosen to be fCFD = 50\%, with the crossing time obtained by linear interpolation between the two samples around the CFD threshold.

\subsection{Track reconstruction and data analysis}
The reconstruction of particle tracks and subsequent data analysis were performed using the Corryvreckan framework \cite{Dannheim_2021}, a modular software package optimized for test beam studies. Data from the six-plane MIMOSA telescope and the FE-I4 reference plane were processed through a standardized Corryvreckan pipeline.

Hits from all detectors were first processed using the EventLoaderEUDAQ2 module. Noisy pixels on the MIMOSA planes, identified by an occupancy exceeding fifty times the average, were masked. Typically, this procedure removed less than 0.1\% of all pixels, corresponding to a loss of below 0.05\% of total hits. Subsequently, the Clustering4D module computed the cluster positions as the geometrical centroids of the constituent hits. The alignment of the telescope and the FE-I4 was performed using the AlignmentTrackChi2 module within the Corryvreckan framework. In this procedure, the MIMOSA planes were aligned by iteratively shifting their coordinates in the \textit{x} and \textit{y} directions and applying small rotations around the beam axis (\textit{z}-axis) relative to a chosen reference plane. The goal was to minimize the residuals between the reconstructed track positions, extrapolated from the other telescope planes, and the measured hit positions in the same MIMOSA plane. This $\chi^2$-based minimization was carried out over approximately $10^5$ events and allowed precise correction of misalignments within the telescope. Accurate alignment is critical for high-precision track reconstruction, which in turn is essential for reliable residual and efficiency measurements at the DUT. Only tracks with a $\chi^2$ per degree of freedom less than six were retained in the analysis. 

Track reconstruction was performed using the Tracking4D module. For data collected at the CERN SPS, track fitting was seeded in the downstream MIMOSA planes. Selected tracks were required to have at least five hits, including one hit in the two MIMOSA planes just before and after the DUT, as well as a matching hit in the FE-I4. Only single-track events were considered. The track fitting procedure at DESY with a 5 GeV electron beam, was slightly modified to account for the differences in experimental setup and beam type. FE-I4 hit information was not available due to data-acquisition synchronization issues, so tracking was performed using only the six MIMOSA planes. For each event, all possible upstream and downstream triplets were reconstructed, and a complete track was defined when a downstream triplet matched an upstream triplet within a minimum distance of approach of 150 $\mathrm{\mu m}$ in the central region. Only events with a single complete track traversing all six MIMOSA planes were retained for analysis.

Following the alignment of the telescope planes, the position and orientation of the DUT were determined relative to the telescope reference frame. The DUT alignment was performed based on the reconstructed track occupancy on the DUT plane. A two-dimensional error-function minimization was applied to the hit occupancy map to extract the optimal $x$, $y$ offsets and in-plane rotation angle $(\theta)$ of the DUT. This procedure corrected for small shifts and rotations of the sensor that can occur between data batches and ensured precise spatial correlation between reconstructed tracks and DUT hits. The resulting geometry file containing the updated $x$, $y$, and $\theta$ parameters was used in subsequent reconstruction steps to define the DUT local coordinates consistently.

The final step of the processing chain was the creation of a merged ROOT file containing both the waveform information recorded by the oscilloscope and the reconstructed track parameters from the telescope system. Event synchronization between the two independent acquisition systems was achieved using timestamp-based matching between the digitizer and the oscilloscope. The difference between the recorded trigger timestamps in both systems was required to be smaller than 0.2 $\mathrm{\mu s}$, corresponding to the clock resolution. The merged dataset served as the input for the subsequent user-level analysis, enabling precise correlation between timing and the position of the hit of the beam particle in the sensor plane on an event-by-event basis. As mentioned earlier, in the offline analysis a cut on the signal amplitude was applied to remove contributions from noise. Table~\ref{tab:selection_criteria} summarizes the selection criteria applied in the analysis.

\begin{table}[htbp]
\centering
\caption{Summary of the selection criteria applied in the analysis.}
\smallskip
\label{tab:selection_criteria}
\begin{tabular}{l|>{\centering\arraybackslash}p{8cm}}
\toprule
\textbf{Selection} & \textbf{Requirement} \\ 
\midrule
Track quality & $\chi^{2}/\mathrm{ndf} < 6$ \\ 
Matching criteria & $|\mathrm{timestamp}_{\text{dig}} - \mathrm{timestamp}_{\text{osc}}| < 0.2~\mu\text{s}$ \\ 
Number of tracks & 1 \\ \
Amplitude cut & $> 40~(10)~\text{mV}$ for unirradiated (irradiated) sensor \\ 
\bottomrule
\end{tabular}
\end{table}

%% file: Sensor_results.tex
\label{sec:results}

The study presents the performance evaluation of LGAD sensors before and after irradiation, with particle beams using the reconstructed position of the tracks. The primary metric studied is the charge collection, time resolution and hit reconstruction efficiency. Performance parameters are studied as a function of bias voltage and incident angle.

\subsection{Collected Charge}
\label{sec:collected_charge}

\begin{figure}[htbp]
	\centering
	\includegraphics[width=0.45\textwidth]{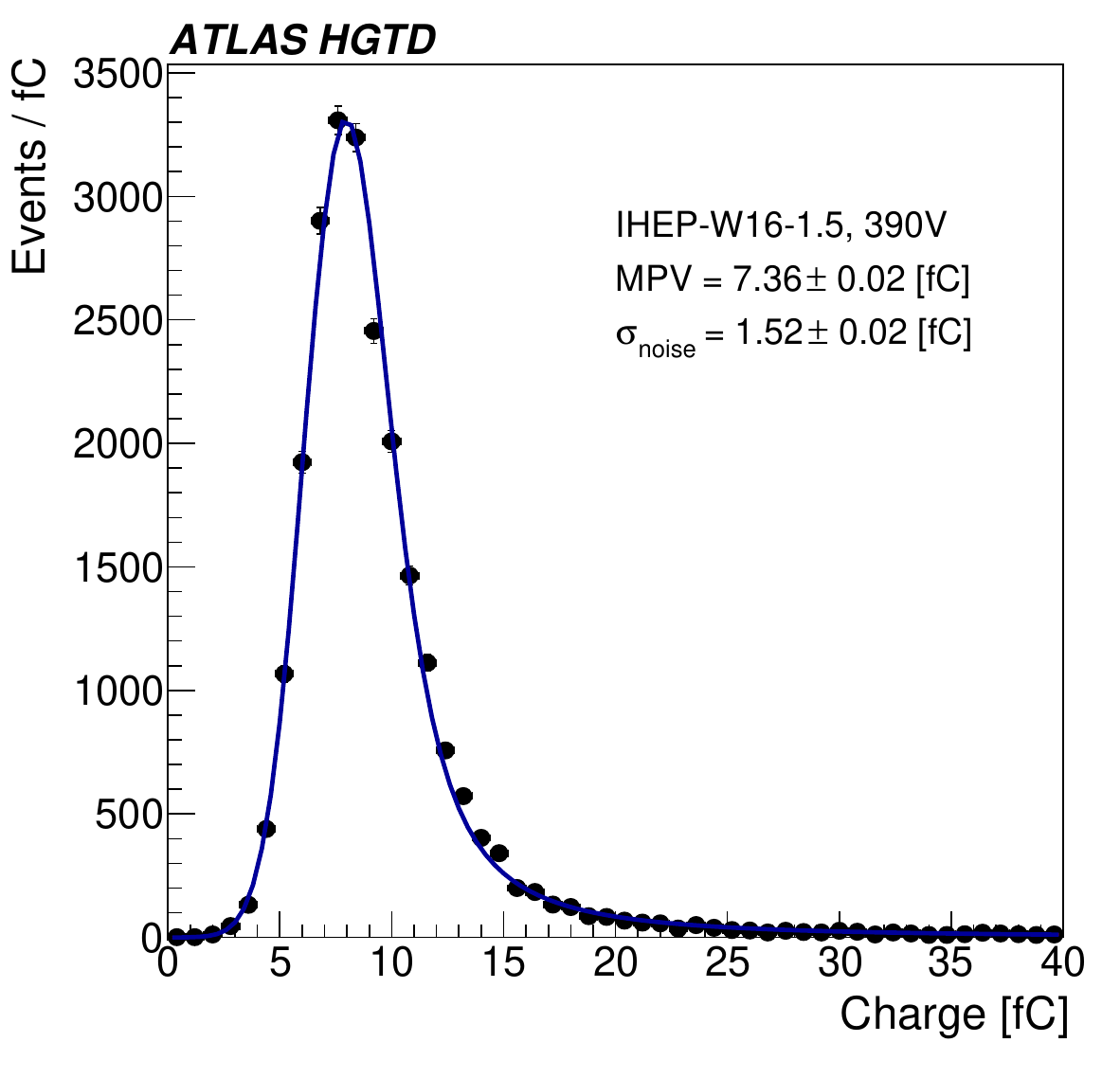}
	
	\caption{Charge distribution for the sensor IHEP-W16-1.5 operated at 390V. The measurement was performed at a temperature of -30$^{\circ}$C. The distribution was fitted with a Landau-Gaussian convoluted function. The collected charge, defined as the most probable value, is 7.36 $\pm$ 0.02 fC. The quoted uncertainty reflects only the statistical component.}
	\label{fig:ihep_charge}
\end{figure}

For each DUT, the charge distribution is obtained from events that passed the selection cuts based on signal amplitude and number of tracks, and is modelled using a Landau-Gaussian convoluted function. The Most Probable Value (MPV) of the fit is used to quantify the collected charge, hereafter referred also as Charge(MPV). Figure~\ref{fig:ihep_charge} shows an example charge distribution for sensor IHEP-W16-1.5. At a bias voltage of 390V the MPV is 7.36 fC. The Gaussian width $\sigma_{\text{noise}}$ = 1.52 indicates a moderate noise contribution, corresponding to a signal-to-noise ratio of about 4.8.

\begin{figure}[htbp]
	\centering
	\subfigure[]{\includegraphics[width=0.45\textwidth]{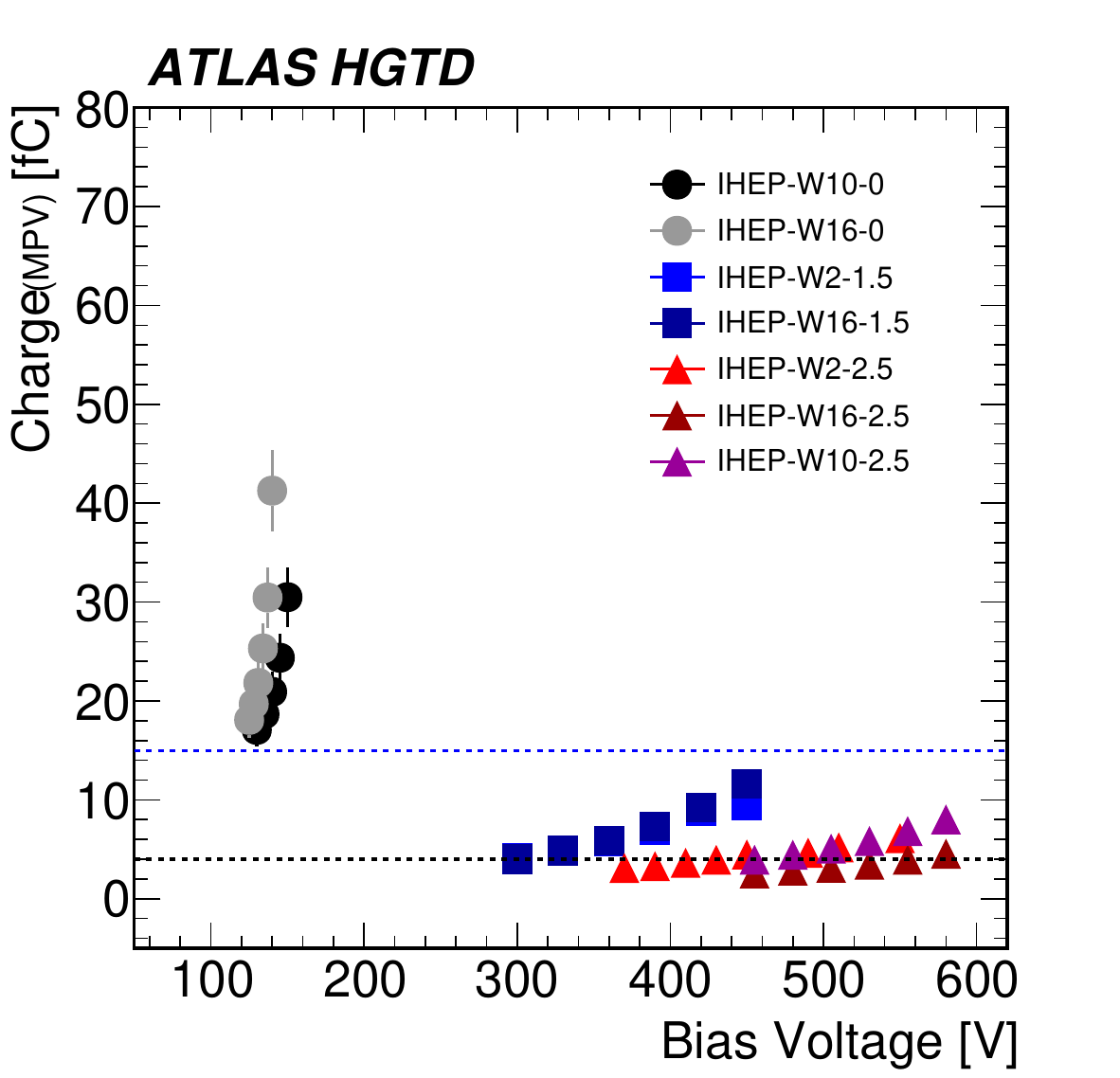}}
	\subfigure[]{\includegraphics[width=0.45\textwidth]{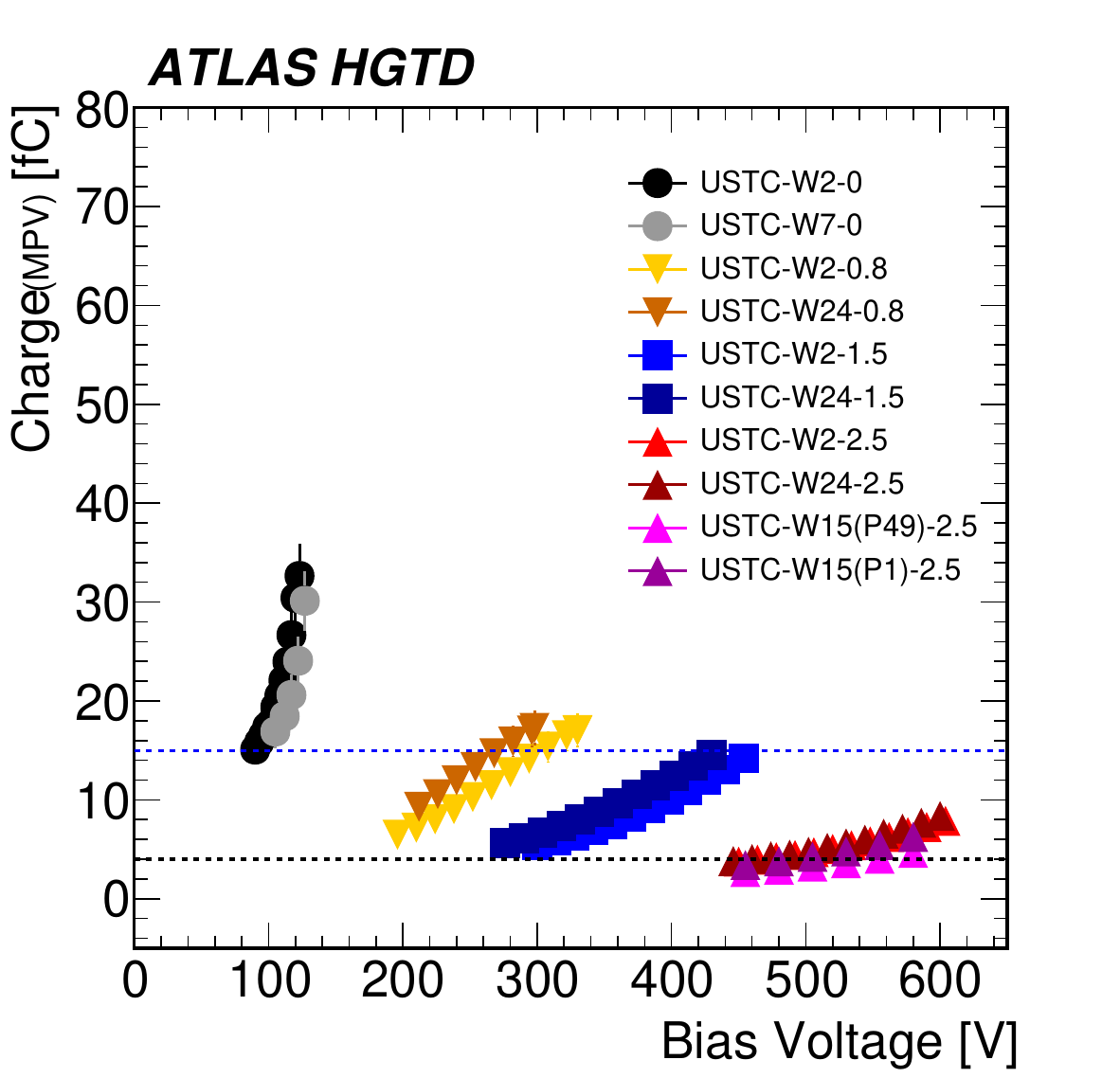}}
	\caption{Collected charge as a function of bias voltage for IHEP (a) and USTC (b) sensors at various irradiation fluences. All measurements were performed at -30$^{\circ}$C, except for the IHEP-W2-1.5 and IHEP-W2-2.5 sensors tested at DESY, where measurements were carried out in the temperature range from -40$^{\circ}$C to -25$^{\circ}$C. The blue (black) dashed line indicates the minimum required charge of 15~fC (4~fC) for unirradiated (irradiated) sensors.}
	\label{fig:charge_collection}
\end{figure}

The MPV is obtained in the same way for different DUTs at various bias voltages. Figure~\ref{fig:charge_collection} presents the collected charge as a function of bias voltage for single-pad sensors irradiated at various fluences, for the IHEP and USTC designs, respectively. The results include measurements for fluences of \(0\), \(0.8\), \(1.5\), and \(2.5 \times 10^{15}~n_{\text{eq}}/\text{cm}^2\). The error bars represent the contribution from the systematic uncertainty assigned to the experimental setup. The blue dashed line indicates the minimum charge collection requirement of 15~fC for unirradiated sensors, while the black dashed line marks the 4 fC threshold requirement for irradiated sensors. Unirradiated sensors consistently meet the 15 fC requirement even at low bias voltages. For irradiated devices, the collected charge is notably reduced due to radiation-induced damage to the gain layer. DUTs exposed to fluences of $0.8 - 2.5 \times 10^{15}~n_{\text{eq}}/\text{cm}^2$ require significantly higher bias voltages to collect the required charge. At the highest fluence, $\phi = 2.5 \times 10^{15}~n_{\text{eq}}/\text{cm}^2$, all sensors pass the 4 fC threshold for a bias voltage $\lesssim$ 550 V, confirming their radiation tolerance up to the HGTD end-of-life scenario.

\subsection{Collected Charge as a function of the incident angle}
\label{subsec:collected_charge_angle}
\begin{figure}[htbp]
	\centering
	\subfigure[]{\includegraphics[width=0.45\textwidth]{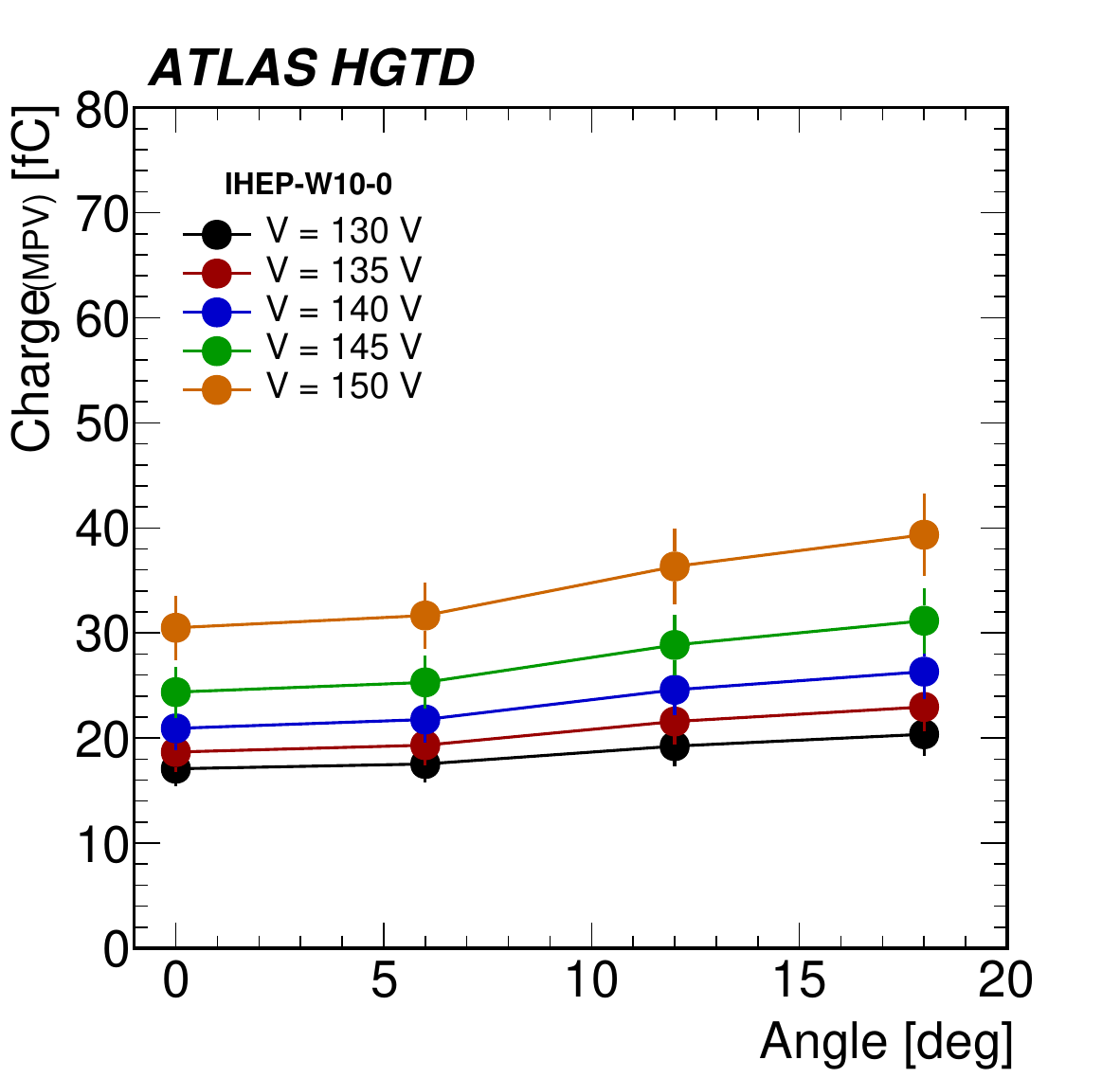}}
	\subfigure[]{\includegraphics[width=0.45\textwidth]{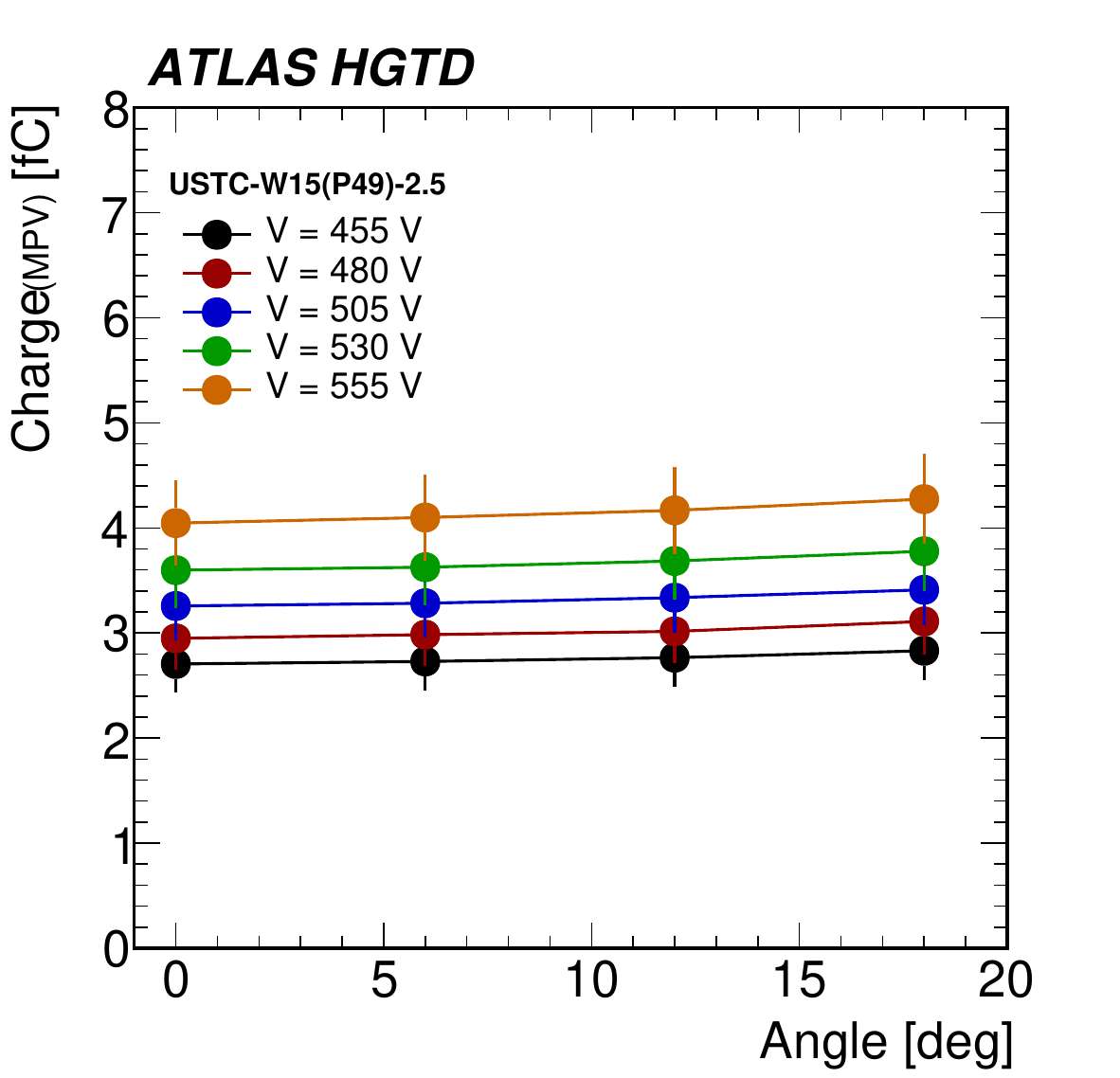}}
	\caption{Collected charge as a function of the angles for IHEP-W10-0 (a) and USTC-W15(P49)-2.5 (b) sensors at various bias voltage points. All measurements were performed at a temperature of -30$^{\circ}$C.}
	\label{fig:charge_collection_vs_angles}
\end{figure}

\begin{figure}[htbp]
	\centering
	\subfigure[]{\includegraphics[width=0.45\textwidth]{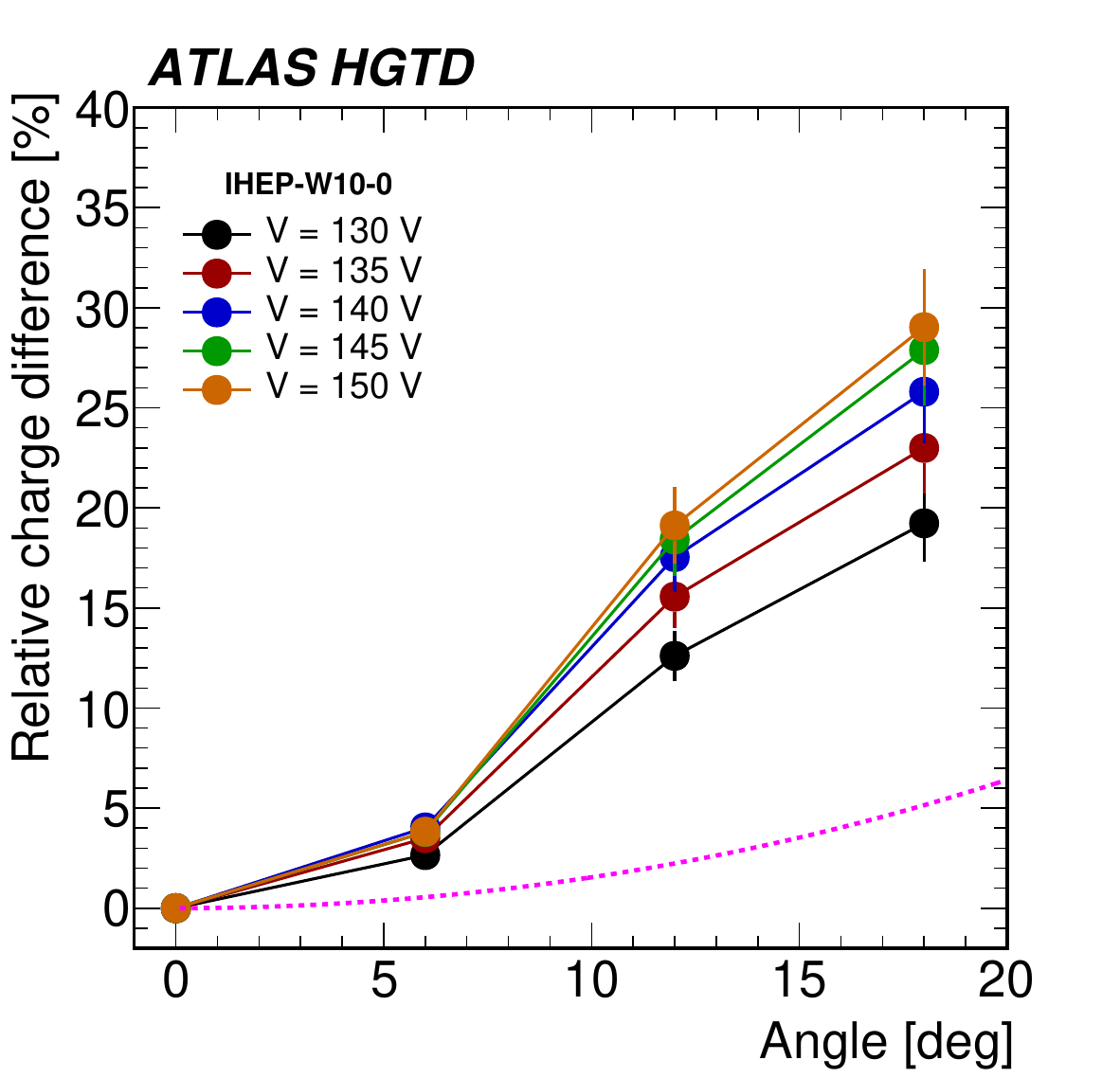}}
	\subfigure[]{\includegraphics[width=0.45\textwidth]{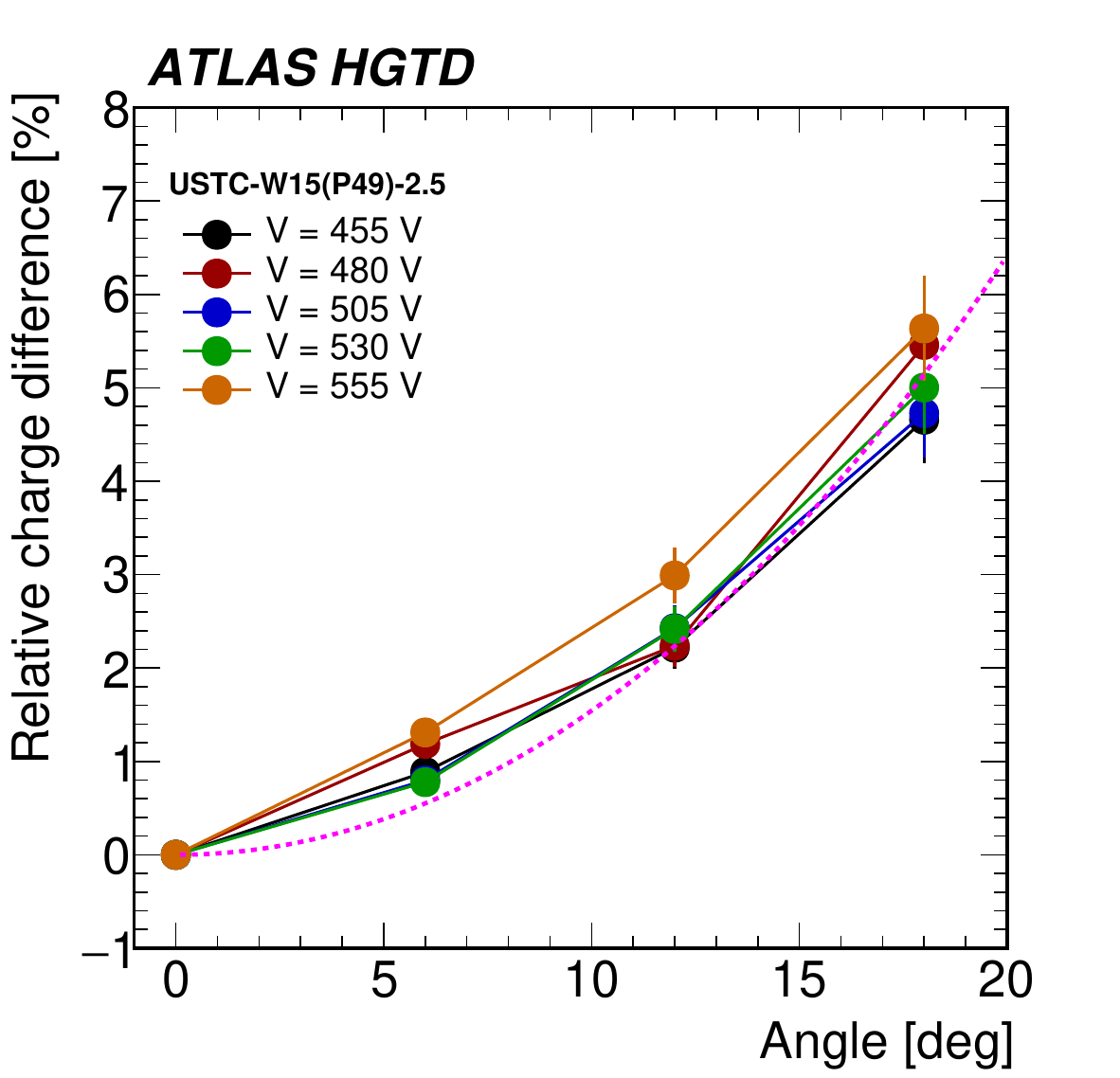}}
	\caption{Relative difference (in \%) of the collected charge with respect to angle 0$^{\circ}$ for IHEP-W10-0 (a) and USTC-W15(P49)-2.5 (b) sensors at various bias voltage points. All measurements were performed at a temperature of -30$^{\circ}$C. The magenta line represents the expected charge increased due to longer path of the charge inside the sensor.}
	\label{fig:Rel_charge_collection_vs_angles}
\end{figure}

The LGAD performance was studied as a function of the angle of the incident beam. Figure~\ref{fig:charge_collection_vs_angles} shows the collected charge as a function of the beam incident angle for an unirradiated IHEP sensor and an irradiated USTC sensor. In both cases, the collected charge increases with the incident angle. This effect is due to electrons drifting towards the gain layer along the projection of the particle track. When the beam is perpendicular to the sensor (0 degrees), the charge density on the gain layer is highly localized and there is a potential charge screening~\cite{Jimenez-Ramos:2022}. Differently, angled tracks produce a more dispersed charge and thus a reduced charge density across the gain layer, making the suppression mechanism less effective. In addition, the collected charge increases also because a longer particle path length through the sensor results in a larger signal amplitude, and therefore a larger collected charge. This effect increases the collected charge by a factor $1/\cos(\alpha)$ where $\alpha$ is the DUT tilted angle. The increase in collected charge with incident angle is even more apparent when examining the relative difference (in \%) of the collected charge with respect to the 0$^{\circ}$ angle, as shown in Figure~\ref{fig:Rel_charge_collection_vs_angles}. Both the screening effect and the particle path inside the sensor are larger for larger value of collected charge, thus the relative increase of charge (MPV) is more significant for unirradiated sensors at higher bias voltage. For irradiated sensors, the relative difference follows the expected trend (magenta line) due to the longer path, demonstrating that the screening effect becomes negligible.

\subsection{Time resolution}
\label{sec:time_resolution}

The time resolution is the key parameters when assessing LGAD sensor performance and it is determined by:
\begin{equation}
\sigma^{2}_{\text{LGAD}} = \sigma^2_{\text{Landau}} + \sigma^2_{\text{Time walk}} + \sigma^2_{\text{Jitter}}\ .
\end{equation}
The Landau term, $\sigma^2_{\text{Landau}}$, arises from non-uniformities in energy deposition along the particle path inside the detector. This contribution depends on the active thickness of the sensor and thin sensors are beneficial. However, thinner sensors suffer from large capacitance and low deposited charge. An active thickness of 50 $\mu$m has been chosen as the best compromise and will be implemented for the LGADs in the HGTD detector.\\
The time walk effect, $\sigma^2_{\text{Time walk}}$, originates from the fact that signals with different amplitudes cross a fixed discriminator threshold at different times. This effect is mitigated in the test beam results by using dedicated time reconstruction techniques, such as the CFD method described in Section~\ref{sec:digitizer_data_processing}. \\
Finally, the jitter term, $\sigma^2_{\text{Jitter}}$, results from the electronic noise and is proportional to the signal time rise and inversely proportional to the signal slope. Minimizing jitter requires a high signal-to-noise ratio and short rise time which is achievable with thin sensors.

\begin{figure}[htbp]
        \centering
        \includegraphics[width=0.45\textwidth]{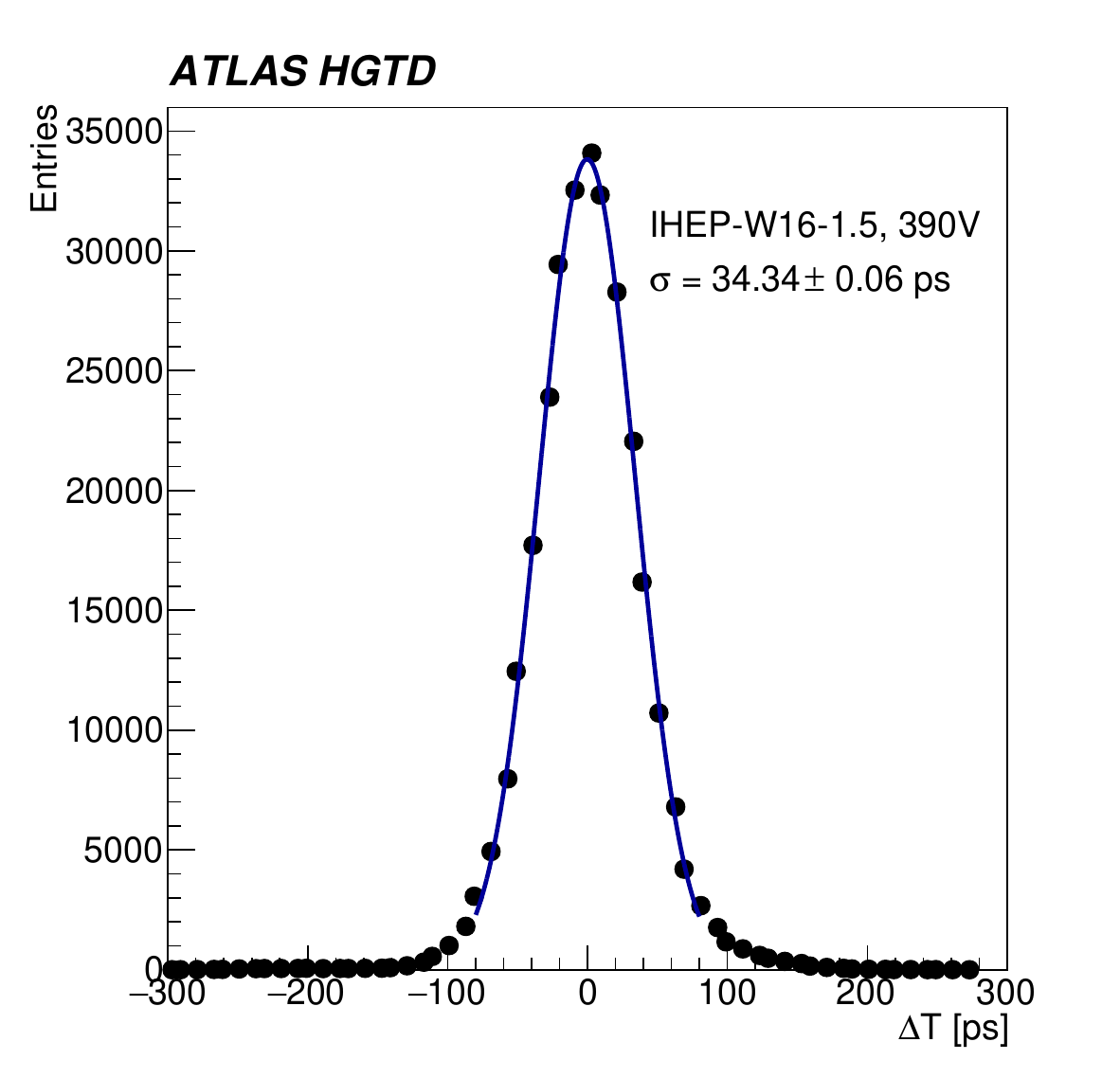}
	\caption{TOA difference between the sensor IHEP-W16-1.5, operated at 390V, and the MCP-PMT. The measurement was performed at a temperature of -30$^{\circ}$C. The distribution is fitted with a Gaussian function, and the width $\sigma$ is extracted from the fit parameters. The quoted uncertainty reflects only the statistical component.}
        \label{fig:dT_IHEP-W16-1.5}
\end{figure}

The time resolution is obtained from the residual distribution between the TOA of the DUT and the MCP-PMT. The width of the residual $\sigma$ is a convolution of the sensor resolution and the resolution of the MCP-PMT:
\begin{equation}
\sigma = \sigma_{\text{DUT}} \oplus \sigma_{\text{MCP-PMT}} \ ,
\end{equation}
where $\sigma_{\text{DUT}}$ is the DUT time resolution and $\sigma_{\text{MCP-PMT}}$ is MCP-PMT time resolution, which is 10.6 $\pm$ 2.2 ps. $\sigma$ is estimated by a Gaussian fit. Figure~\ref{fig:dT_IHEP-W16-1.5} shows an example distribution of the TOA difference between the IHEP-W16-1.5 sensor and the MCP-PMT. The distribution shows non-Gaussian tails due to noise. The time resolution of the DUT is obtained by subtracting the MCP-PMT contribution from the measured residual width:
\begin{equation}
\sigma_{\text{DUT}} = \sqrt{\sigma^2 - \sigma_{\text{MCP-PMT}}^2}
\end{equation} 

\begin{figure}[htbp]
	\centering
	\subfigure[\label{fig:time_res_voltage_USTC_unirradiated}]{\includegraphics[width=0.45\textwidth]{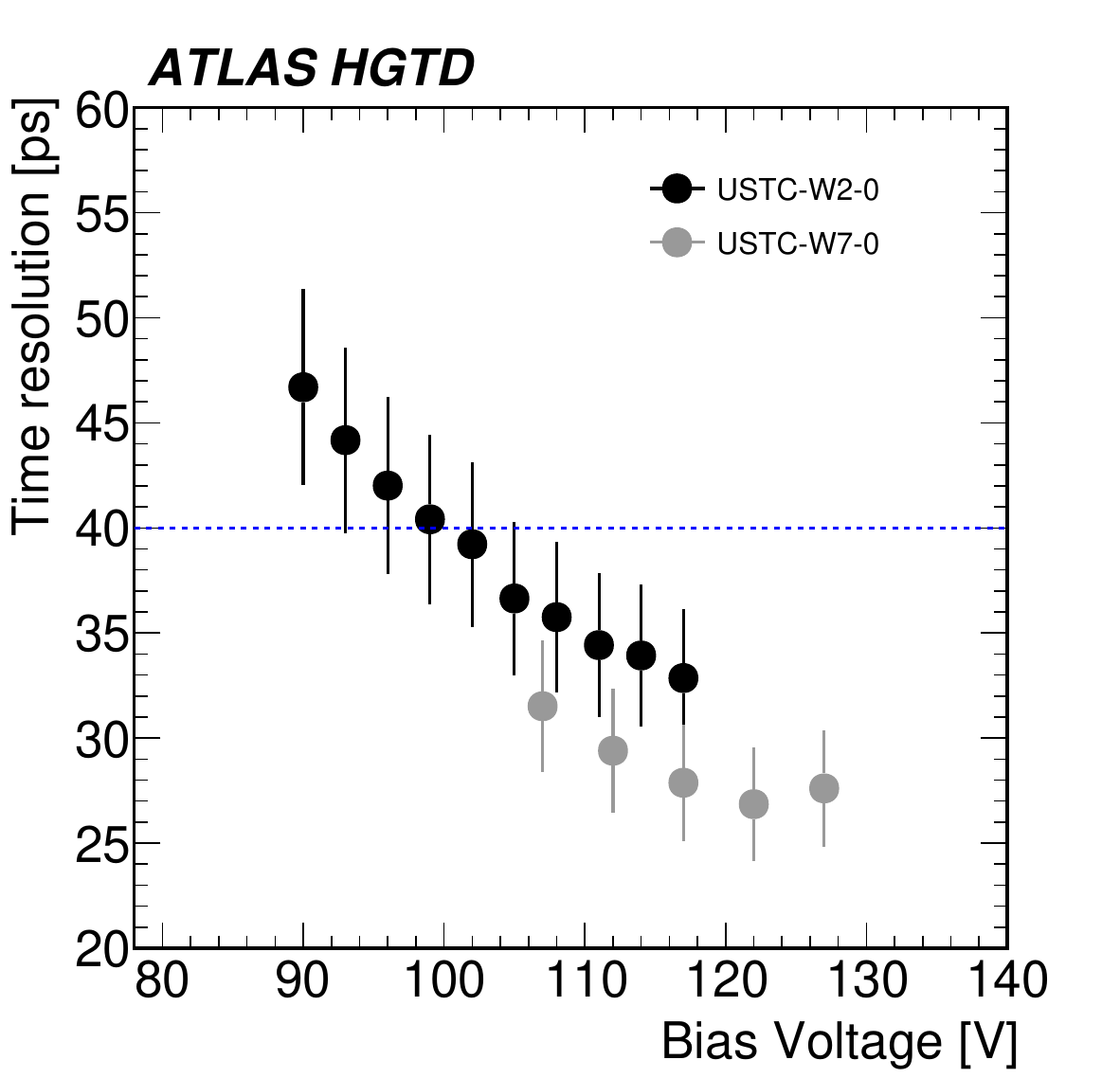}	}
	\subfigure[\label{fig:time_res_voltage_USTC_irradiated}]{\includegraphics[width=0.45\textwidth]{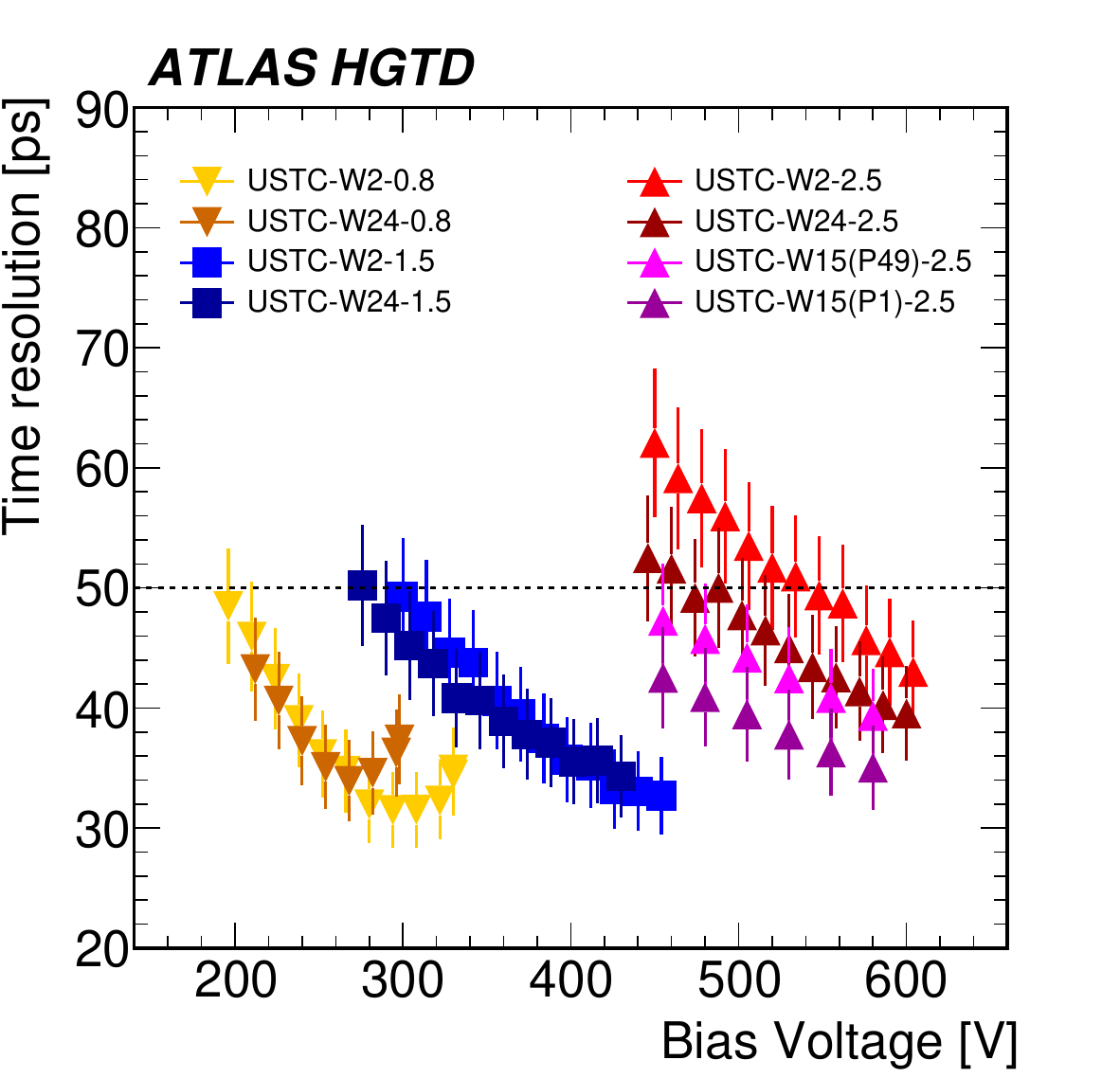}}
	\caption{Time resolution as a function of the bias voltage for USTC unirradiated (a) and irradiated (b) sensors. All measurements were performed at a temperature of -30$^{\circ}$C. The blue (black) dashed line at 40 ps (50 ps) represents the sensor-only per-hit time resolution requirements for unirradiated (irradiated) sensors.}
	\label{fig:time_res_voltage_USTC}
\end{figure}

\begin{figure}[htbp]
	\centering
	\subfigure[\label{fig:time_res_voltage_IHEP_unirradiated}]{\includegraphics[width=0.45\textwidth]{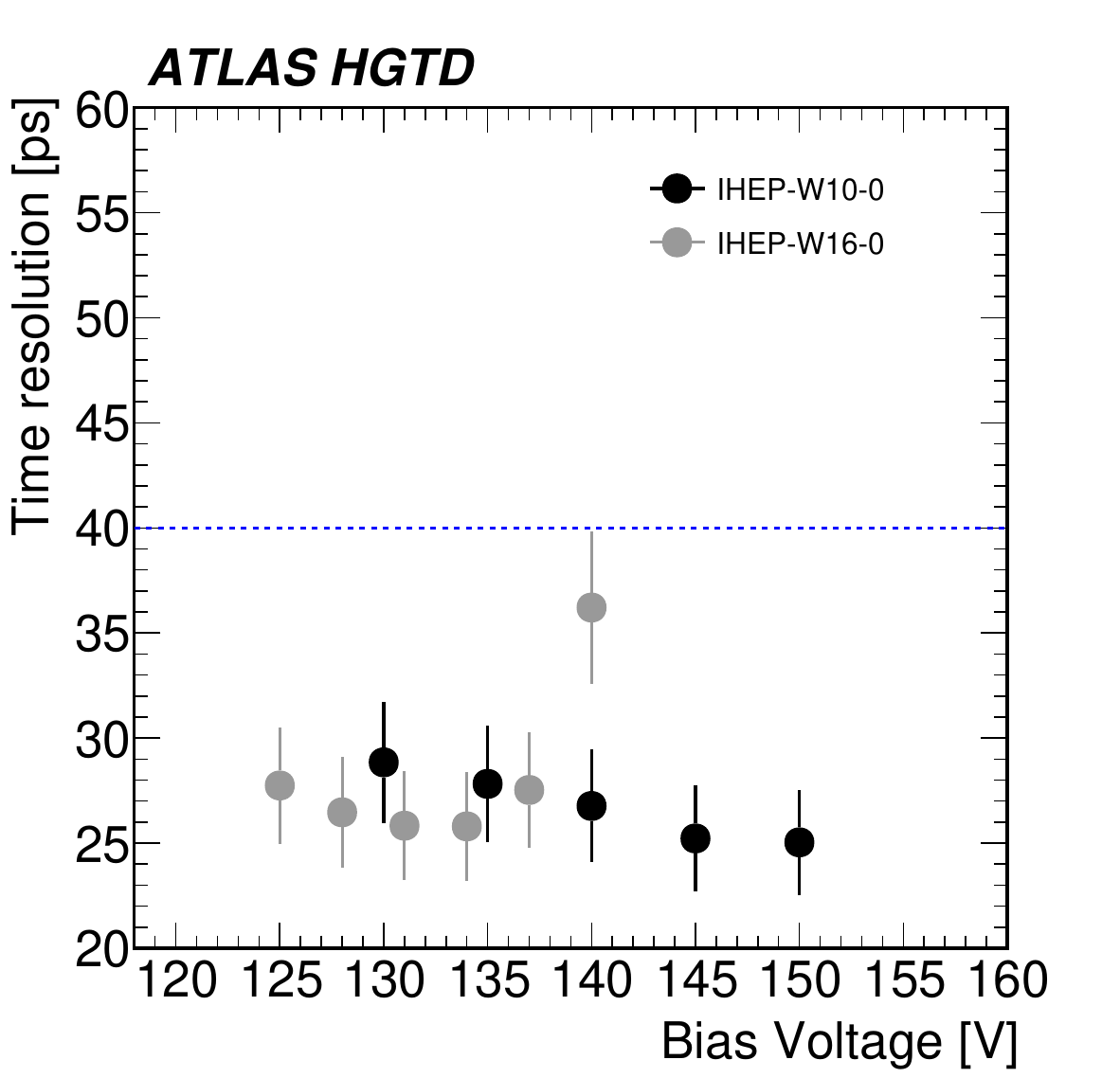}}
	\subfigure[\label{fig:time_res_voltage_IHEP_irradiated}]{\includegraphics[width=0.45\textwidth]{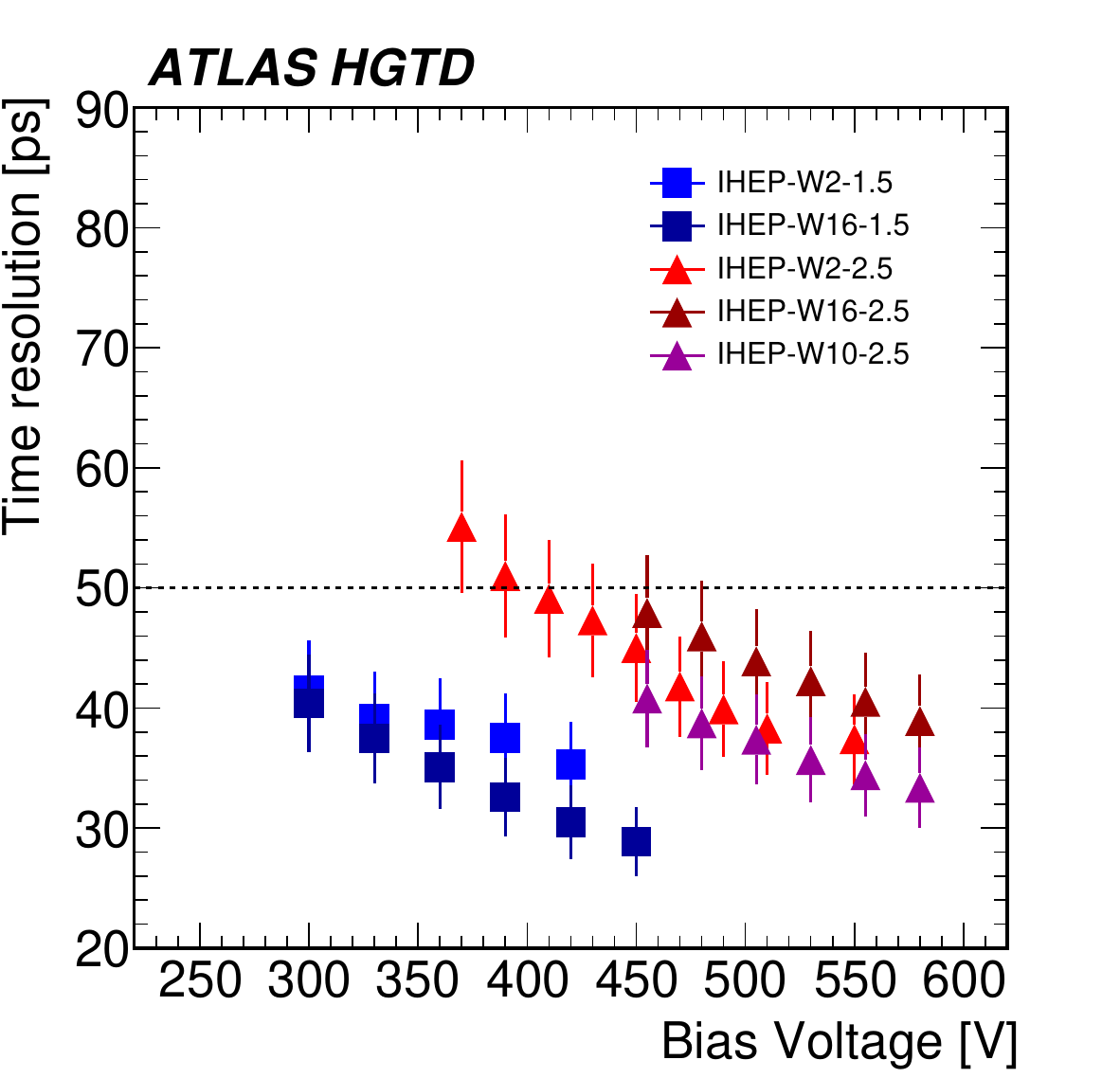}}
	\caption{Time resolution as a function of the bias voltage for IHEP unirradiated (a) and irradiated (b) sensors. All measurements were performed at -30$^{\circ}$C, except for the IHEP-W2-1.5 and IHEP-W2-2.5 sensors tested at DESY, where measurements were carried out in the temperature range from -40$^{\circ}$C to -25$^{\circ}$C. The blue (black) dashed line at 40 ps (50 ps) represents the sensor-only per-hit time resolution requirements for unirradiated (irradiated) sensors.}
	\label{fig:time_res_voltage_IHEP}
\end{figure}

Figures~\ref{fig:time_res_voltage_USTC} and ~\ref{fig:time_res_voltage_IHEP} show the time resolution as a function of the bias voltage for the USTC and IHEP sensors, respectively. Due to differences in the bias voltage settings for sensors with different irradiation levels, the plots are separated into unirradiated and irradiated cases. As before, the error bars represent the contribution from the systematic uncertainty assigned to the experimental setup. 
For the least irradiated USTC sensors (yellow and brown points on Fig.~\ref{fig:time_res_voltage_USTC_irradiated}) and the unirradiated IHEP sensors (grey points on Fig.~\ref{fig:time_res_voltage_IHEP_unirradiated}) the time resolution worsens at higher bias voltages. This degradation is attributed to noise hits, which become significant when operating the sensors at bias voltages close to the breakdown voltage. 
Time resolution differences among unirradiated sensors can arise from variations in charge-carrier velocity saturation, which depends on the electric field \cite{article_electronDrift}, and thus on the sensor bias.
All sensors meet the requirement of a time resolution below 40 ps for unirradiated sensors and below 50 ps for irradiated sensors.

\subsection{Time resolution as a function of the incident angle}
\label{subsec:time_angle}

\begin{figure}[htbp]
	\centering
	\subfigure[]{\includegraphics[width=0.45\textwidth]{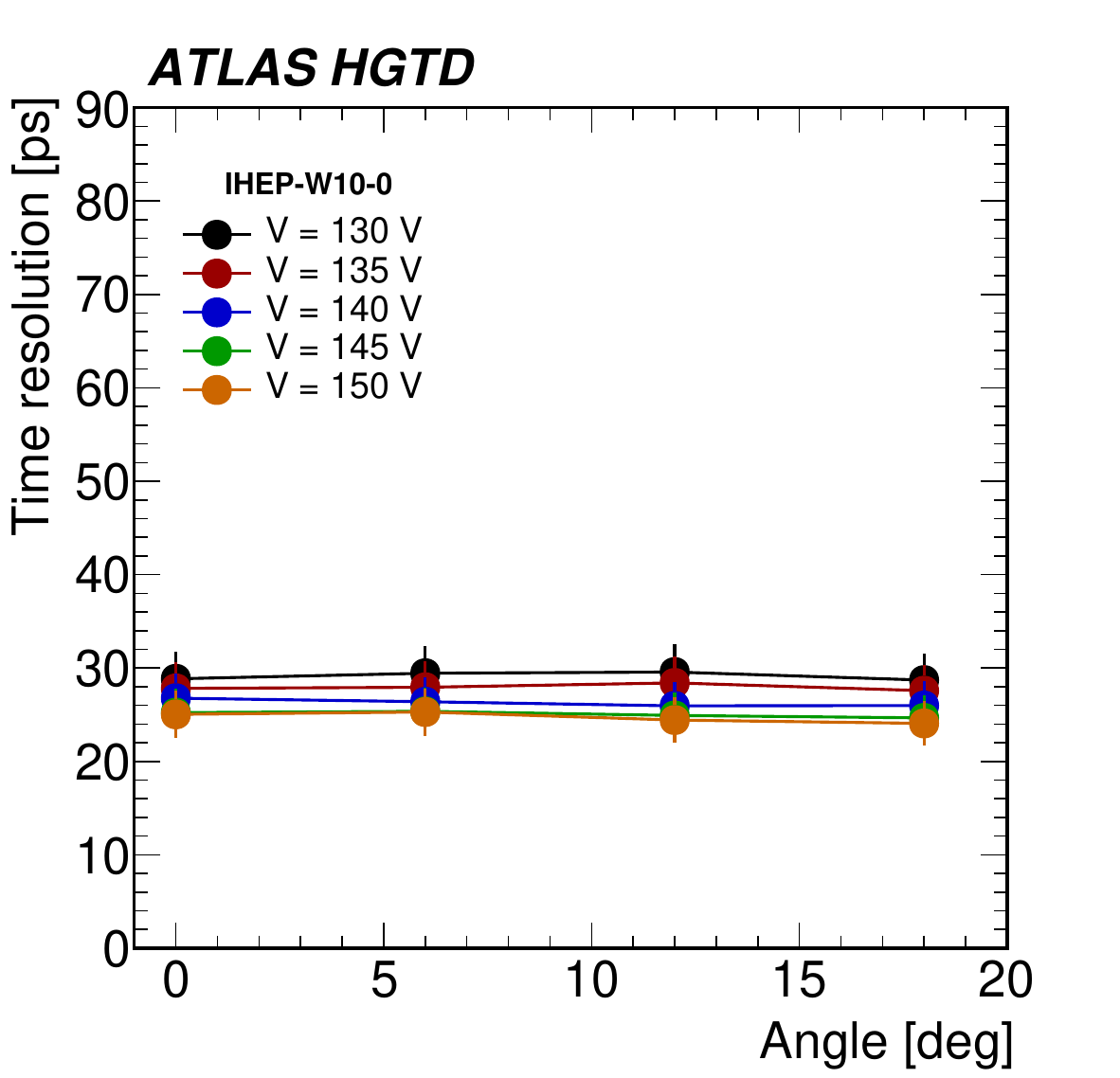}}
	\subfigure[]{\includegraphics[width=0.45\textwidth]{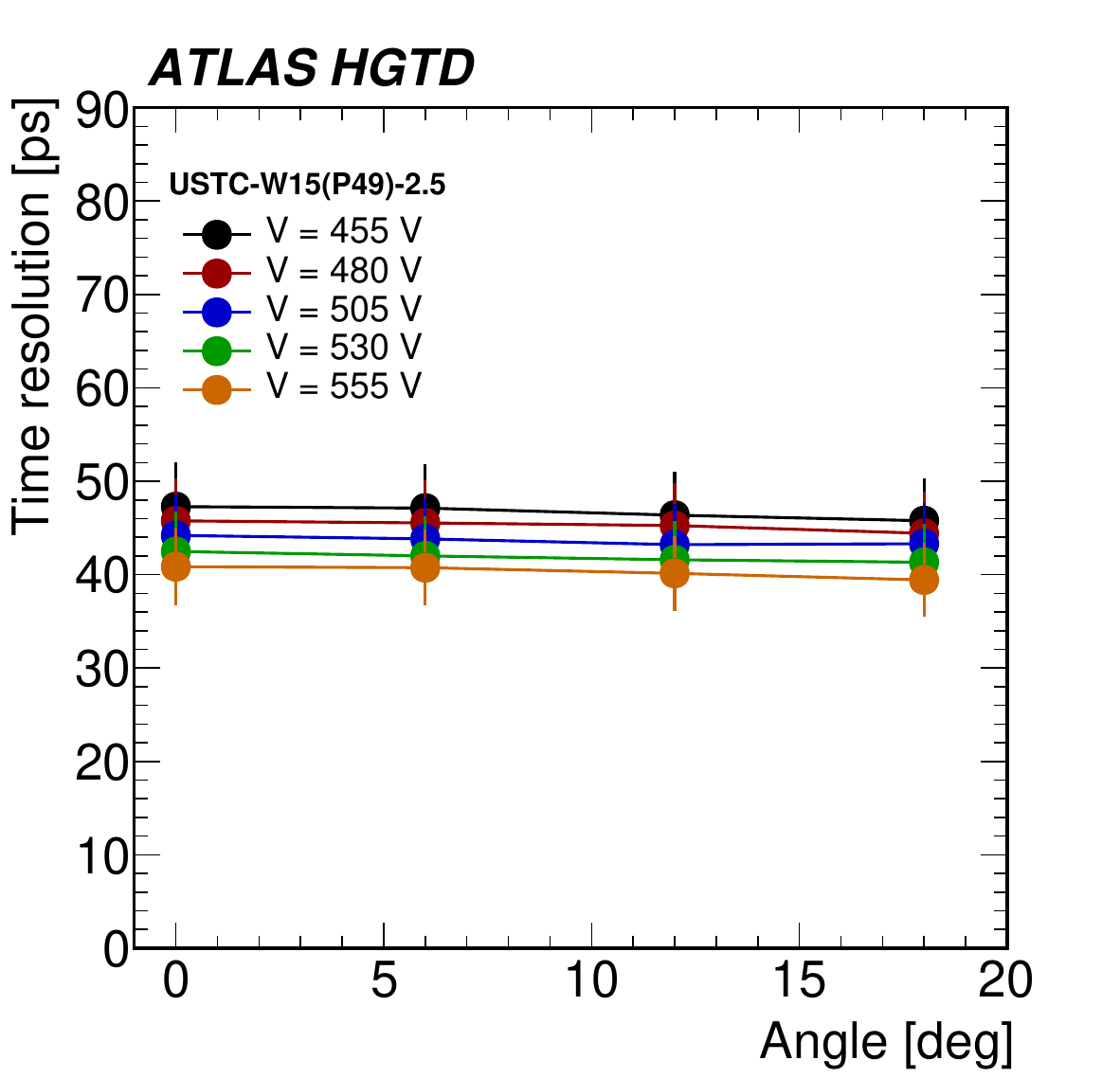}}
	\caption{Time resolution as a function of the angles for IHEP-W10-0 (a) and USTC-W15(P49)-2.5 (b) sensors at various bias voltage points. All measurements were performed at a temperature of -30$^{\circ}$C.}
	\label{fig:timeResolution_vs_angles}
\end{figure}

\begin{figure}[htbp]
	\centering
	\subfigure[]{\includegraphics[width=0.45\textwidth]{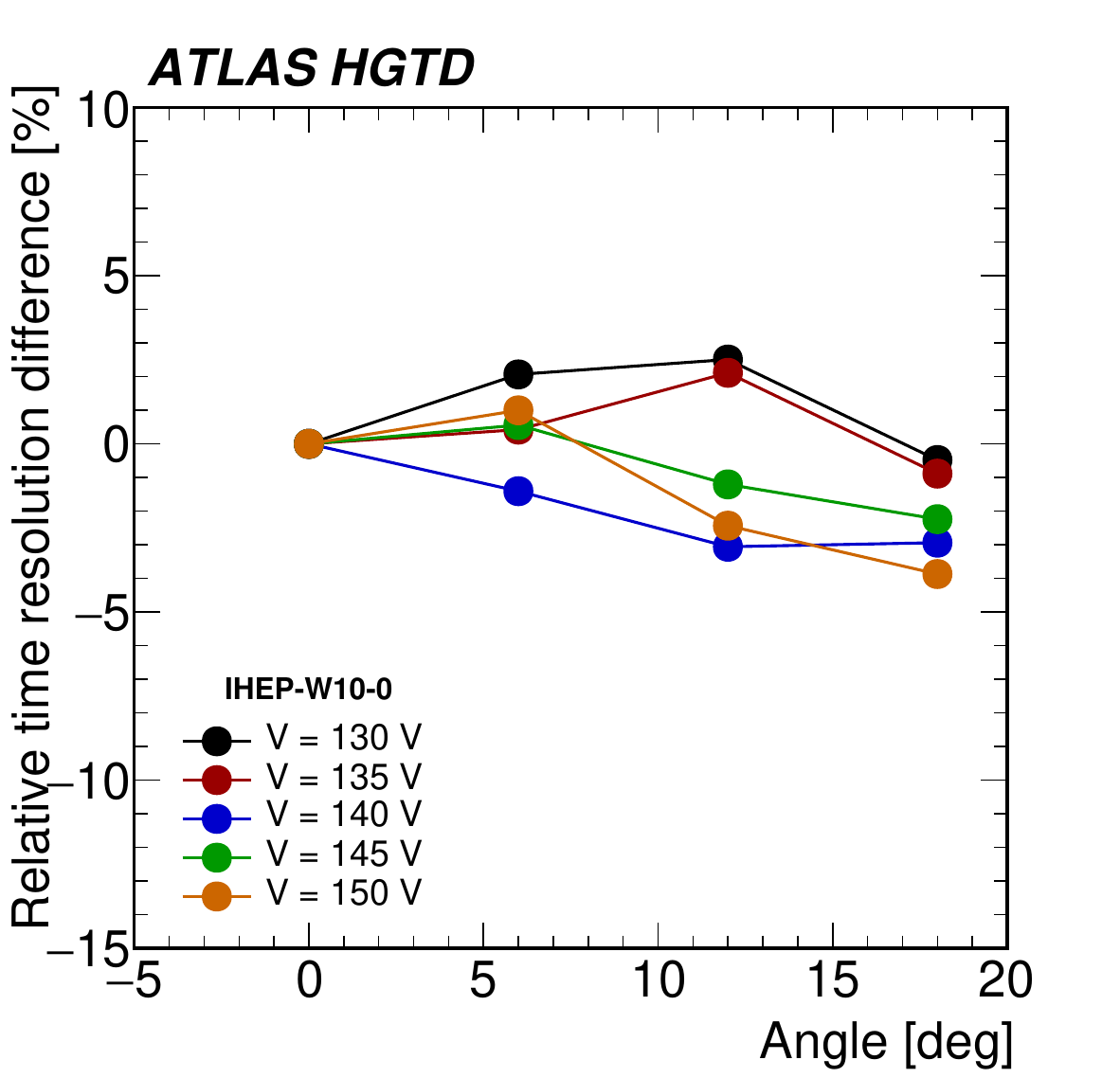}}
	\subfigure[]{\includegraphics[width=0.45\textwidth]{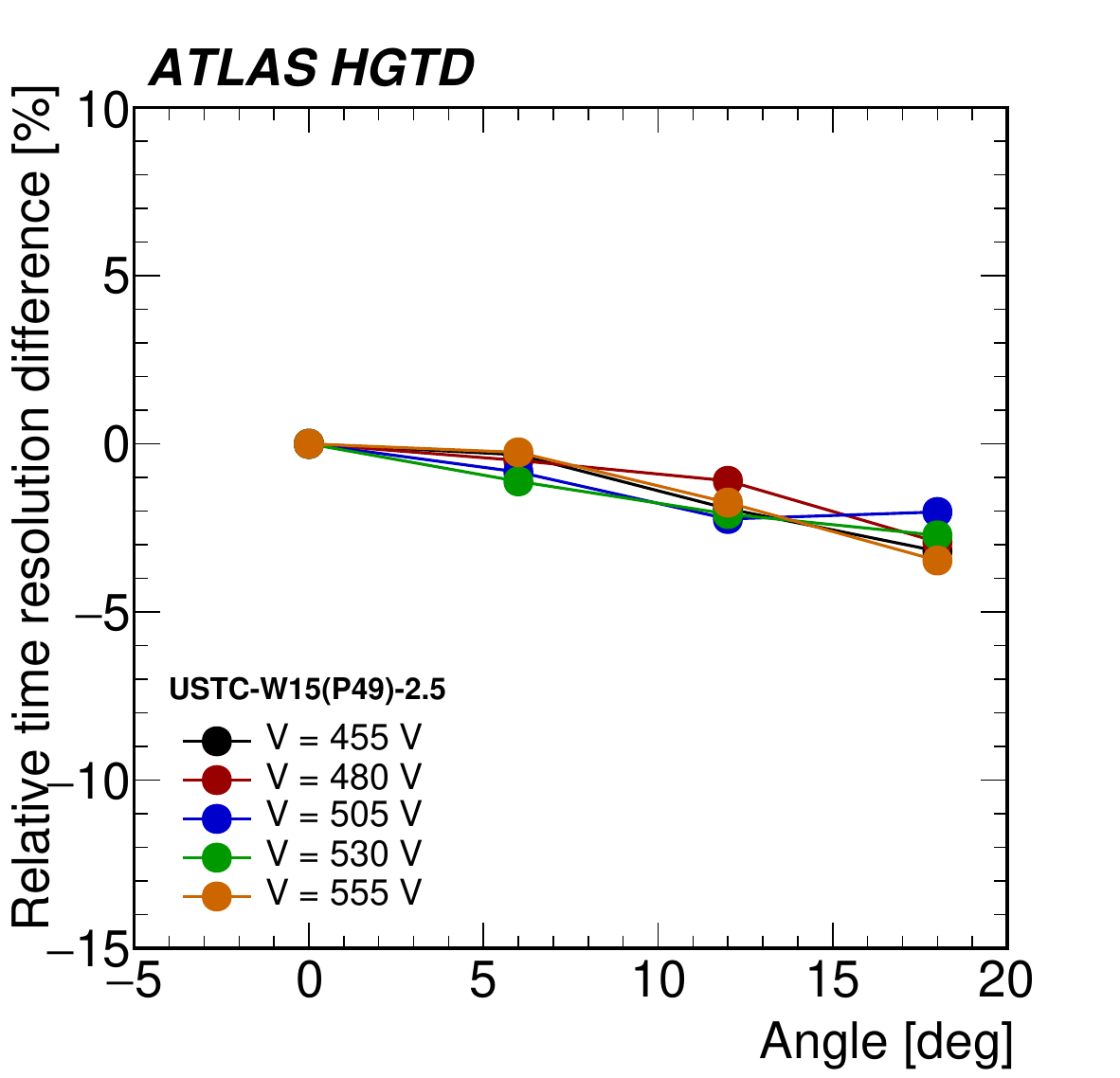}}
	\caption{Relative difference (in \%) of the time resolution with respect to angle 0$^{\circ}$ for IHEP-W10-0 (a) and USTC-W15(P49)-2.5 (b) sensors at various bias voltage points. All measurements were performed at a temperature of -30$^{\circ}$C.}
	\label{fig:Rel_time_vs_angles}
\end{figure}

The time resolution has also been studied as a function of the rotation of the sensor with respect to the beam. Figure~\ref{fig:timeResolution_vs_angles} shows the time resolution as a function of the beam incident angle for an unirradiated IHEP sensor and an irradiated USTC sensor. The time resolution shows little dependence on the incident angle. This is confirmed by the relative variation in time resolution with respect the incident angle at 0$^{\circ}$, as shown in Figure~\ref{fig:Rel_time_vs_angles}. The relative differences in time resolution remain within a few percent.


 
\subsection{Hit Reconstruction Efficiency}
\label{sec:efficiency}

The hit reconstruction efficiency is defined as the ratio of reconstructed tracks producing a signal above threshold in the sensor fiducial region over the total number of reconstructed tracks crossing that region:
\begin{equation}
	\text{Hit Efficiency} = \frac{\text{Reconstructed tracks with } q > Q_{\text{cut}}}{\text{Total reconstructed tracks}} \ 
\end{equation}
where $Q_{\text{cut}}$ is the minimum collected charge, set to 2~fC. The fiducial region is defined as a rectangle covering the central 60\% of the DUT area\footnote{The central area is evaluated considering -0.81 (-0.87) mm < \textit{x} < 0.81 (0.87) mm and -0.31 (-0.29) mm < \textit{y} < 0.31 (0.29) mm for USTC (IHEP) sensors.}, excluding edge effects, where the sensor response is expected to be uniform. 
The hit reconstruction efficiency is calculated only for events with exactly one reconstructed track to discard possible pile-up events. The efficiency study has been performed only using data collected at CERN. 

\begin{figure}[htbp]
	\centering
	\includegraphics[width=0.5\textwidth]{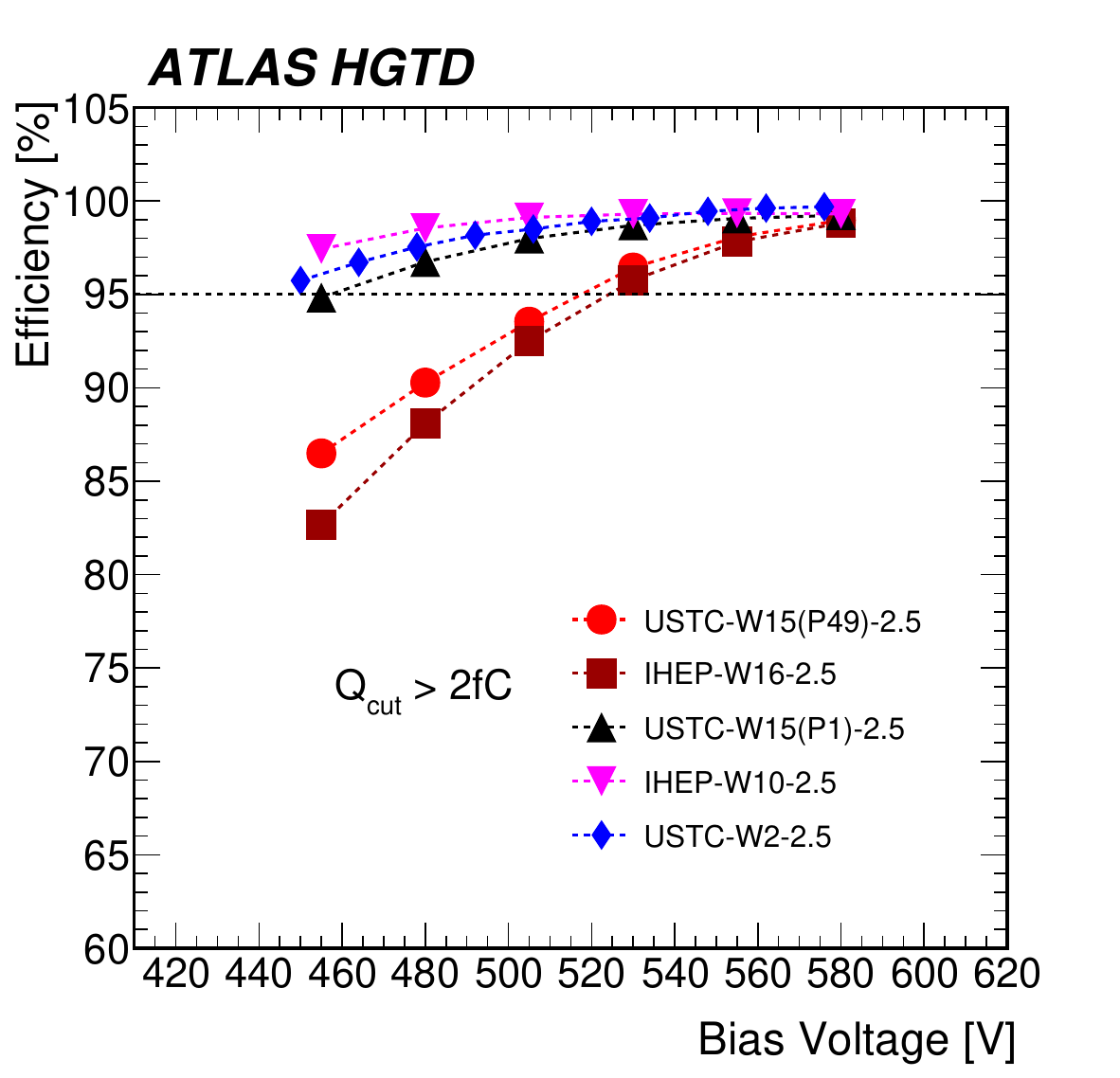}
	\caption{Hit reconstruction efficiency versus bias voltage for sensors irradiated at $2.5 \times 10^{15}~n_{\text{eq}}/\text{cm}^2$. All measurements were performed at a temperature of -30$^{\circ}$C. Results are shown for multiple sensor designs from USTC and IHEP. The dashed line indicates the 95\% efficiency requirement for HGTD operation. A charge threshold of $Q_{\text{cut}} >2~\mathrm{fC}$ is applied.}
	\label{fig:efficiency}
\end{figure}

For unirradiated sensors ($\phi = 0 \times 10^{15}~n_{\text{eq}}/\text{cm}^2$) and sensors with intermediate irradiation level ($\phi = 0.8 - 1.5 \times 10^{15}~n_{\text{eq}}/\text{cm}^2$), the hit reconstruction efficiency remains above 99\% across the entire tested bias voltage range. 
Figure~\ref{fig:efficiency} presents the efficiency as a function of bias voltage for the most irradiated sensors  ( $\phi = 2.5 \times 10^{15}~n_{\text{eq}}/\text{cm}^2$). 
The results demonstrate that all sensor designs achieve the minimum required efficiency of 95\% for HGTD operation in the end of life scenario, though at different bias voltages. 

\begin{figure}[htbp]
	\centering
	\subfigure[]{\includegraphics[width=0.45\textwidth]{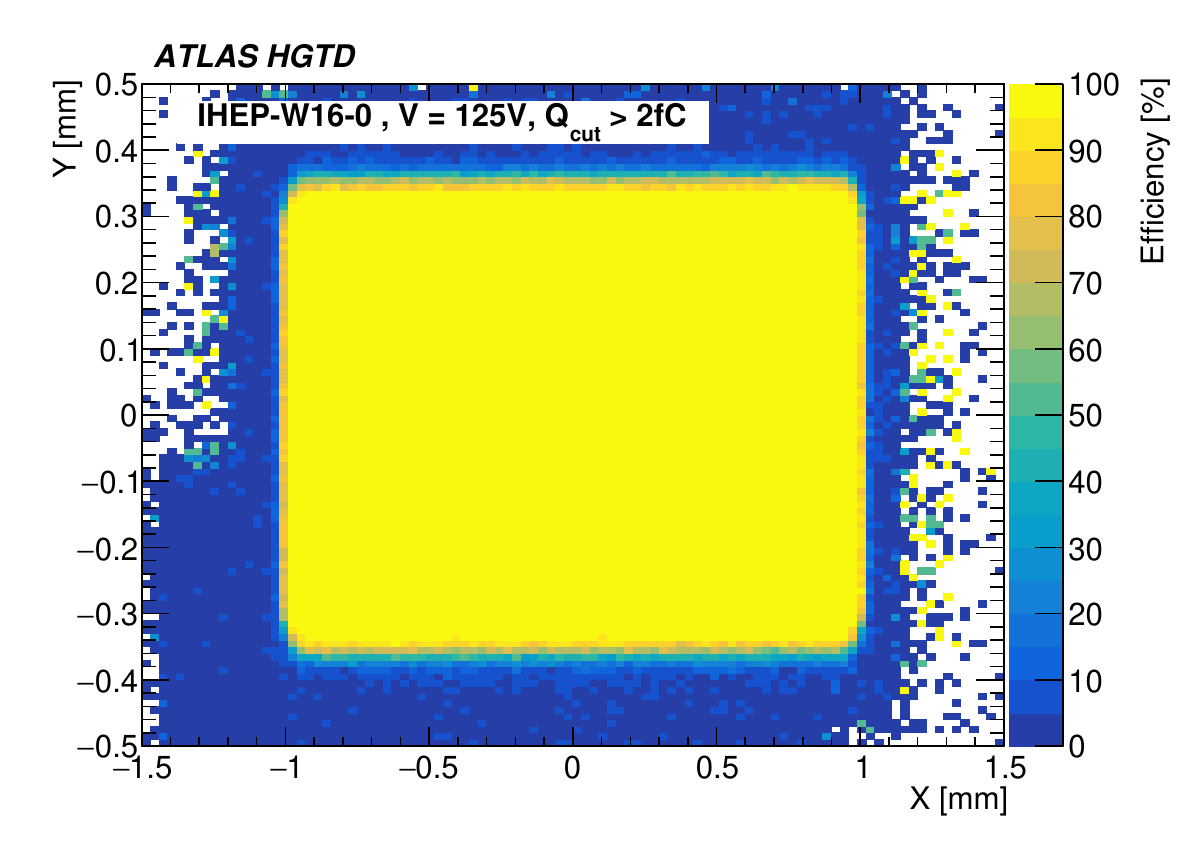}}
	\subfigure[]{\includegraphics[width=0.45\textwidth]{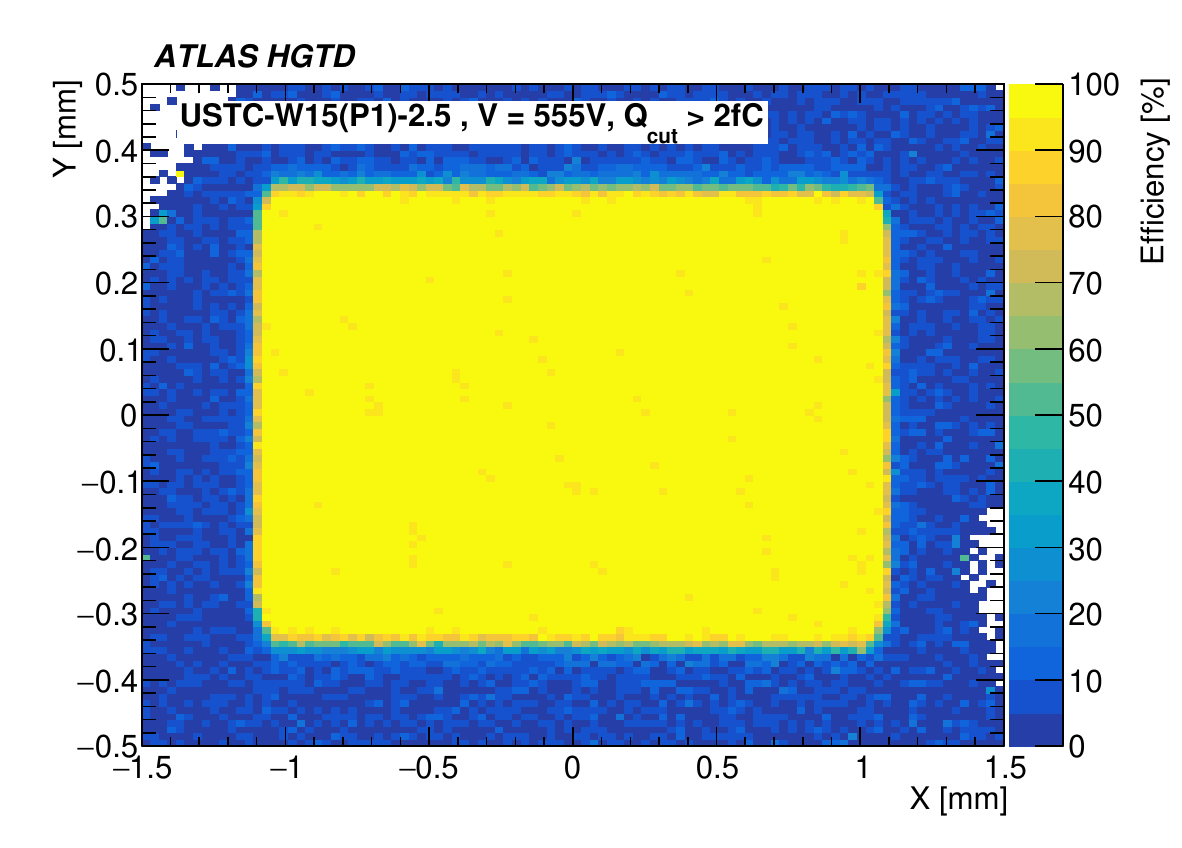}}
	\caption{2D maps of the efficiency as a function of the hit position in the sensor plane for IHEP-W16-0 sensor operated at a bias voltage of 125 V (a) and for the USTC-W15(P1)-2.5 sensor operated at a bias voltage of 555 V (c). All measurements were performed at a temperature of -30$^{\circ}$C.}
	\label{fig:2Deff_maps}
\end{figure}

To study the sensor response uniformity, the efficiency was determined as a function of the hit position of the track on the sensor. Figure~\ref{fig:2Deff_maps} shows the 2D efficiency maps as a function of the hit position for sensors irradiated to different fluences. The plots demonstrate a high degree of uniformity across the entire sensor area, regardless of the sensor irradiation.

\begin{figure}[htbp]
	\centering
	\subfigure[]{\includegraphics[width=0.45\textwidth]{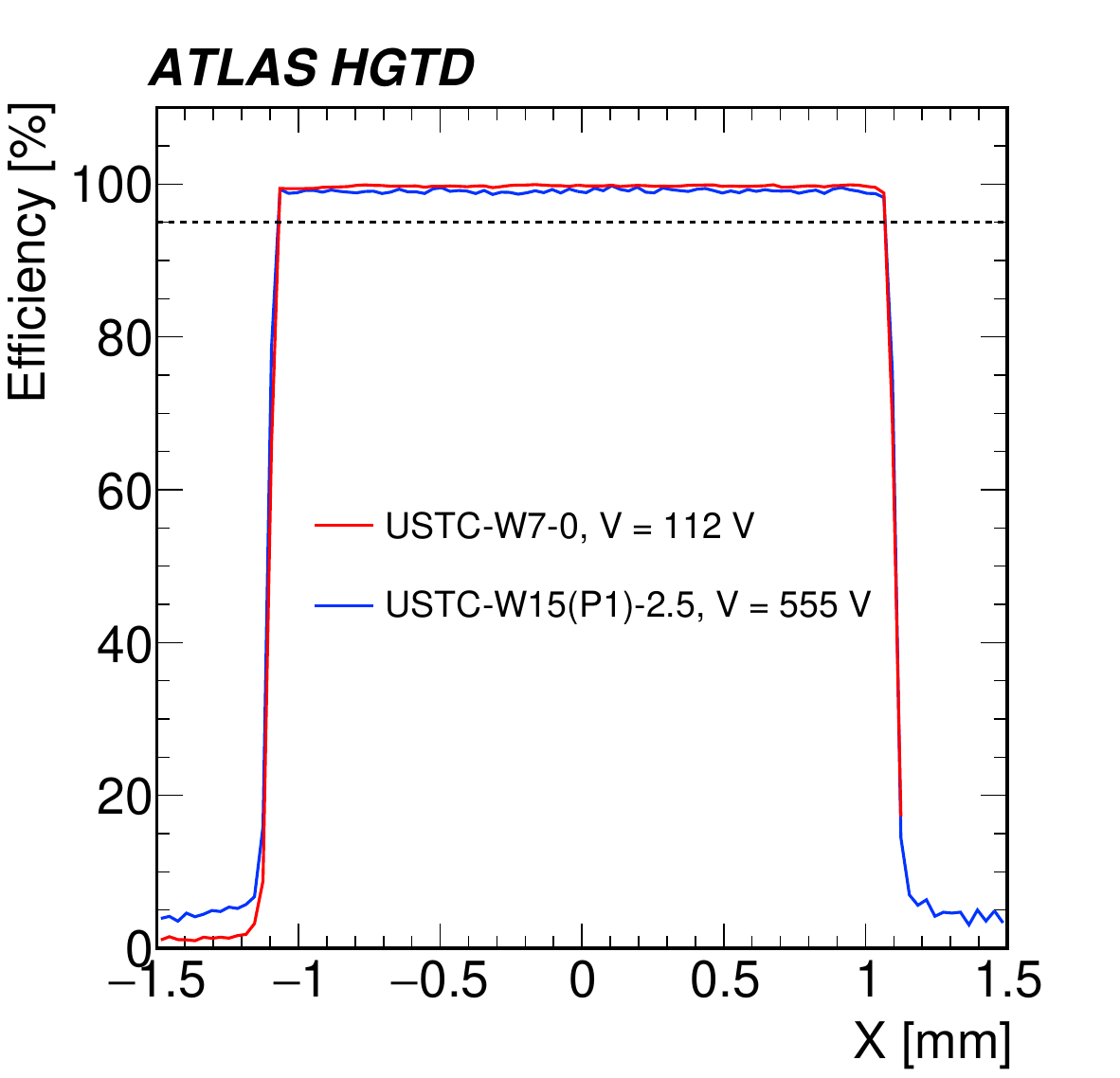}}
	\subfigure[]{\includegraphics[width=0.45\textwidth]{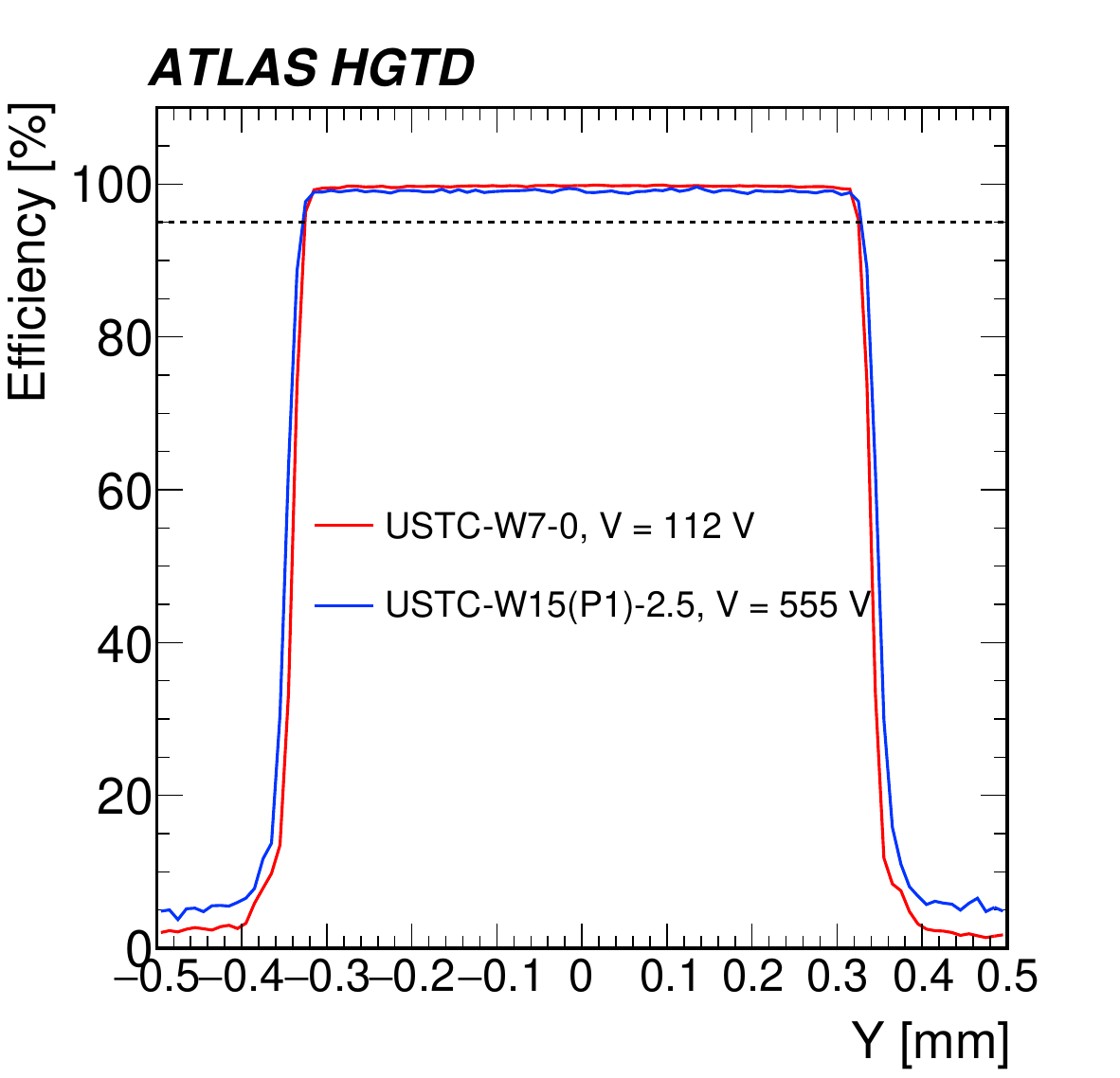}}\\
	\subfigure[]{\includegraphics[width=0.45\textwidth]{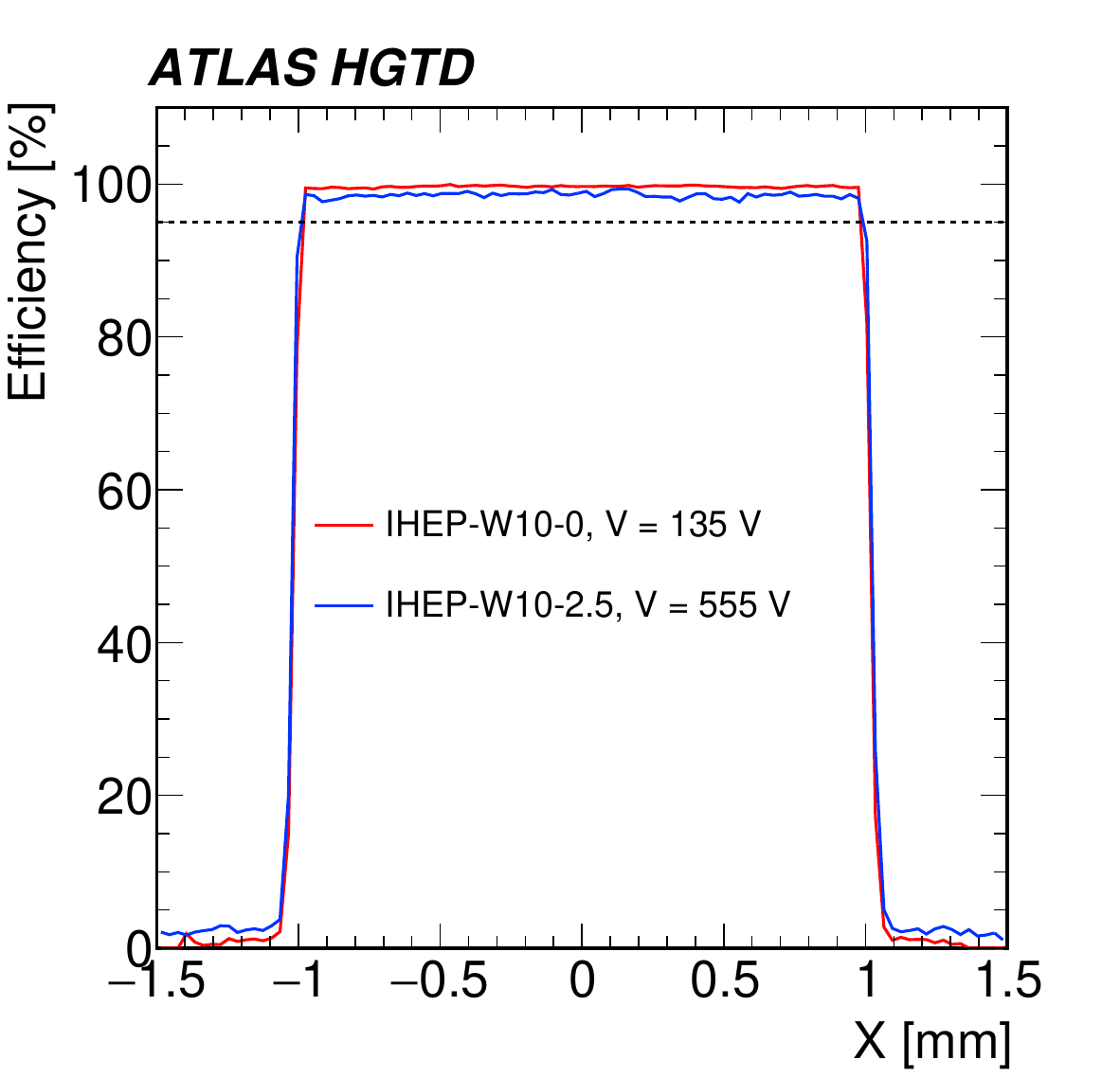}}
	\subfigure[]{\includegraphics[width=0.45\textwidth]{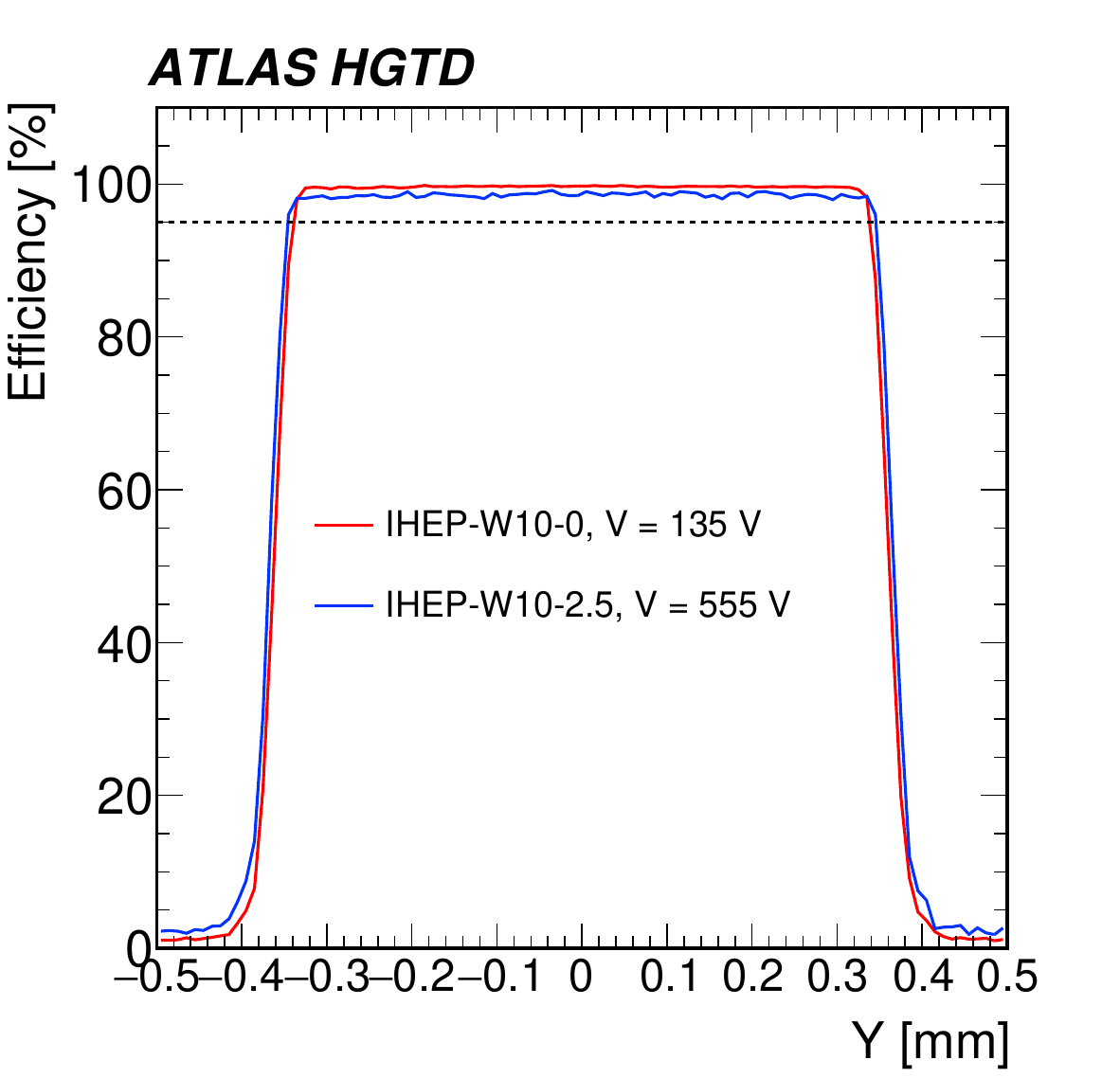}} 
	\caption{Projections on the \textit{x}- and \textit{y}-axis of the efficiency in the central region of the sensor for USTC (a,b) and IHEP (c,d) sensors. All measurements were performed at a temperature of -30$^{\circ}$C. Each plot shows the comparison of the projection of the efficiency for unirradiated (red line) and irradiated sensors (blue line). For USTC-W7-0, the right tail of the distribution is truncated to avoid spikes due to the missing statistics. The projections on the \textit{x}-axis (\textit{y}-axis) are evaluated considering the integral of the central region from -0.31 mm to 0.31 mm (from -0.81 mm to 0.81 mm) and from -0.29 mm to 0.29 mm (from -0.87 mm to 0.87 mm) for USTC and IHEP sensors, respectively.}
	\label{fig:eff_projY}
\end{figure}

Figure~\ref{fig:eff_projY} shows the uniformity of the efficiency along the \textit{x}-axis and \textit{y}-axis for an unirradiated sensor and a sensor irradiated to a fluence of $2.5 \times 10^{15}~n_{\text{eq}}/\text{cm}^2$ for each of the two designs, respectively. In both cases, the efficiency within the active area remains uniform and exceeds 95\%. Same conclusion is found for studying the characteristics of the other sensors. A comparison between unirradiated and irradiated devices reveals an apparent increase in the active area after irradiation. This effect is not due to physical expansion, but rather to radiation-induced modifications of the electric-field distribution and charge-collection dynamics. As a result, the sensor edges become less sharp, effectively enlarging the active region.
The sensor size can be estimated at the point where the efficiency reaches 50\% and the results are summarised in Table \ref{tab:sensors_size}. These numbers should be compared to the nominal dimensions reported in Tab.~\ref{tab:sensors_list}. Table \ref{tab:sensors_size} also gives the size of the region where the efficiency is larger than 95\% and 97\%, which is the required efficiency in the 60\% central area for irradiated and unirradiated sensors, respectively.

%
\begin{table}[h]
\centering
\caption{Size of unirradiated and irradiated sensors for both sensor designs. 
The sensor size is estimated from the region where the efficiency is larger 
than 50\%, 95\% and 97\%. The applied bias voltage for each measurement is also shown.}
\smallskip
\label{tab:sensors_size}
\footnotesize
\begin{adjustbox}{width=\textwidth}
\begin{tabular}{lccccccc}
\toprule
\textbf{Device name} &
\makecell{\textbf{Bias voltage} \\ \textbf{[V]}} &
\makecell{\textbf{Size on \textit{x}-axis} \\ \textbf{at 50\% [mm]}} &
\makecell{\textbf{Size on \textit{x}-axis} \\ \textbf{at 95\% [mm]}} &
\makecell{\textbf{Size on \textit{x}-axis} \\ \textbf{at 97\% [mm]}} &
\makecell{\textbf{Size on \textit{y}-axis} \\ \textbf{at 50\% [mm]}} &
\makecell{\textbf{Size on \textit{y}-axis} \\ \textbf{at 95\% [mm]}} &
\makecell{\textbf{Size on \textit{y}-axis} \\ \textbf{at 97\% [mm]}} \\
\midrule
USTC-W7-0          & 112 & 2.20 $\pm$ 0.01 & 2.15 $\pm$ 0.01 & 2.14 $\pm$ 0.01 & 0.68 $\pm$ 0.01 & 0.65 $\pm$ 0.01 & 0.65 $\pm$ 0.01 \\
USTC-W15(P1)-2.5   & 555 & 2.21 $\pm$ 0.01 & 2.15 $\pm$ 0.01 & 2.14 $\pm$ 0.01 & 0.70 $\pm$ 0.01 & 0.66 $\pm$ 0.01 & 0.65 $\pm$ 0.01 \\
\midrule
IHEP-W10-0         & 135 & 2.04 $\pm$ 0.01 & 1.97 $\pm$ 0.01 & 1.96 $\pm$ 0.01 & 0.72 $\pm$ 0.01 & 0.68 $\pm$ 0.01 & 0.67 $\pm$ 0.01 \\
IHEP-W10-2.5       & 555 & 2.05 $\pm$ 0.01 & 2.00 $\pm$ 0.01 & 1.99 $\pm$ 0.01 & 0.74 $\pm$ 0.01 & 0.69 $\pm$ 0.01 & 0.69 $\pm$ 0.01 \\
\bottomrule
\end{tabular}
\end{adjustbox}
\end{table}

%% file: Conclusions.tex
\label{sec:conclusion}

This study presents a comprehensive evaluation of the performance of single LGADs from the HGTD pre-production. The LGADs developed by IHEP and USTC were tested under varying irradiation conditions, bias voltages, and beam incident angles, with a focus on key parameters relevant for HGTD operation at the HL-LHC.
 
Unirradiated sensors consistently meet the 15~$\mathrm{fC}$ requirement even at relatively low bias voltages, confirming their suitability for high-precision timing applications. As expected, radiation-induced damage significantly reduces the collected charge for irradiated devices. Nevertheless, for fluences up to $\phi = 2.5 \times 10^{15}~\mathrm{n_{eq}/cm^2}$,  all tested sensors provided a collected charge above 4~$\mathrm{fC}$ within the prescribed operating bias voltage, demonstrating radiation tolerance compatible with HGTD requirements. A clear enhancement of collected charge with increasing incident angle is observed for unirradiated sensors, attributed to the more favorable geometry of charge deposition and reduced charge screening effects. 

Time resolution improves with increasing bias voltage for most irradiated sensors, reaching below 50~$\mathrm{ps}$ even at the highest tested fluences. All sensors meet the requirement of a time resolution below 40~$\mathrm{ps}$ for unirradiated sensors and below 50~$\mathrm{ps}$ for irradiated sensors. For unirradiated sensors and sensors with moderate irradiation, the hit reconstruction efficiency for $Q_{\text{cut}} > 2~\mathrm{fC}$, remains above 99\% across the entire tested bias voltage range, while all sensors, including those irradiated up to $\phi = 2.5 \times 10^{15}~\mathrm{n_{eq}/cm^2}$, meet the minimum 95\% efficiency requirement. Spatial uniformity studies further confirm consistent performance across the full active area of the sensor, indicating no localized degradation due to irradiation.

These results confirm that both IHEP and USTC LGAD sensors are suitable candidates for the HGTD project, having achieved the required performance in charge collection, time resolution, and hit efficiency even under high radiation fluences. Further measurements with full-size sensors hybridized to the final readout ASICs, as well as comprehensive system-level tests, are ongoing to evaluate the feasibility of the complete LGAD-based timing detector for the HL-LHC.

%% file: acknowledgments.tex
The authors gratefully acknowledge CERN and the SPS staff for successfully operating the North Experimental Area and for continuous supports to the users. Part of the measurements leading to these results have been performed at the Test Beam Facility at DESY Hamburg (Germany), a member of the Helmholtz Association (HGF). We gratefully acknowledge the financial support from the Slovenian Research and Innovation Agency (ARIS P1-0135, ARIS Z1-50011, ARIS J7-4419), Slovenia; the National Natural Science Foundation of China (No. 11961141014, No. 12188102) and the Ministry of Science and Technology of China (No. 2023YFA1605901), China; and FAPESP (2020/04867-2, 2022/14150-3, 2023/18486-9 and 2023/18484-6), MCTI/CNPq (INCT CERN Brasil 406672/2022-9) and CAPES - Code 001, Brazil. We acknowledge the support from BMFTR, Germany; NWO, Netherlands; and FCT, Portugal.

%% file: ANA-HGTD-2025-01-INT1.bib
@techreport{2091129,
  author        = {ATLAS Collaboration},
  title         = {Technical Design Report: A High-Granularity Timing Detector for the ATLAS Phase-II Upgrade},
  institution   = {CERN},
  reportNumber  = {CERN-LHCC-2020-007, ATLAS-TDR-031},
  year          = {2020},
  month         = {06},
  url           = {https://cds.cern.ch/record/2719855},
}

@article{ATLAS:2008xda,
    author = "{ATLAS Collaboration}",
    title = "{The ATLAS Experiment at the CERN Large Hadron Collider}",
    doi = "10.1088/1748-0221/3/08/S08003",
    journal = "JINST",
    volume = "3",
    pages = "S08003",
    year = "2008"
}

@article{Apollinari:2017lan,
  author       = {Apollinari, G. and B\'ejar Alonso, I. and Br{\"u}ning, O. and Fessia, P. and Lamont, M. and Rossi, L. and Tavian, L.},
  title        = "{High-Luminosity Large Hadron Collider (HL-LHC): Technical Design Report V. 0.1}",
  reportNumber = {CERN-2017-007-M},
  doi          = {10.23731/CYRM-2017-004},
  journal      = {CERN Yellow Reports: Monographs},
  volume       = {4/2017},
  year         = {2017}
}

@article{Pellegrini:2014lki,
    author = "Pellegrini, G. and others",
    editor = "Unno, Yoshinobu and Fukazawa, Yasushi and Hou, Suen and Ohsugi, Takashi and Sadrozinski, Hartmut F. -W.",
    title = "{Technology developments and first measurements of Low Gain Avalanche Detectors (LGAD) for high energy physics applications}",
    doi = "10.1016/j.nima.2014.06.008",
    journal = "Nucl. Instrum. Meth. A",
    volume = "765",
    pages = "12--16",
    year = "2014"
}

@article{deLaTaille:2018jyb,
    author = "de La Taille, Christophe and Callier, St\'ephane and Di Lorenzo, Selma Conforti and Seguin-Moreau, Nathalie and Dinaucourt, P and Martin-Chassard, G and Agapopoulou, C and Makovec, N and Serin, L and Simion, S",
    title = "{ALTIROC0, a 20 pico-second time resolution ASIC for the ATLAS High Granularity Timing Detector (HGTD)}",
    reportNumber = "AIDA-2020-CONF-2018-005",
    doi = "10.22323/1.313.0006",
    journal = "PoS",
    volume = "TWEPP-17",
    pages = "006",
    year = "2018"
}

@article{Allaire:2018bof,
    author = "Allaire, C. and others",
    title = "{Beam test measurements of Low Gain Avalanche Detector single pads and arrays for the ATLAS High Granularity Timing Detector}",
    eprint = "1804.00622",
    archivePrefix = "arXiv",
    primaryClass = "physics.ins-det",
    doi = "10.1088/1748-0221/13/06/P06017",
    journal = "JINST",
    volume = "13",
    number = "06",
    pages = "P06017",
    year = "2018"
}

@article{Agapopoulou:2020gnf,
    author = "Agapopoulou, C. and others",
    title = "{Performance of a Front End prototype ASIC for picosecond precision time measurements with LGAD sensors}",
    eprint = "2002.06089",
    archivePrefix = "arXiv",
    primaryClass = "physics.ins-det",
    doi = "10.1088/1748-0221/15/07/P07007",
    journal = "JINST",
    volume = "15",
    number = "07",
    pages = "P07007",
    year = "2020"
}

@article{Lange:2017pxs,
    author = "Lange, J. and others",
    title = "{Gain and time resolution of 45 $\mu$m thin Low Gain Avalanche Detectors before and after irradiation up to a fluence of $10^{15}$ n$_{eq}$/cm$^2$}",
    eprint = "1703.09004",
    archivePrefix = "arXiv",
    primaryClass = "physics.ins-det",
    reportNumber = "AIDA-2020-PUB-2018-007",
    doi = "10.1088/1748-0221/12/05/P05003",
    journal = "JINST",
    volume = "12",
    number = "05",
    pages = "P05003",
    year = "2017"
}

@article{Agapopoulou:2022vxk,
    author = "Agapopoulou, C. and others",
    title = "{Performance in beam tests of irradiated Low Gain Avalanche Detectors for the ATLAS High Granularity Timing Detector}",
    doi = "10.1088/1748-0221/17/09/P09026",
    journal = "JINST",
    volume = "17",
    number = "09",
    pages = "P09026",
    year = "2022"
}

@article{Garcia:2021ujp,
    author = "Garc\'ia, Luc\'ia Castillo and Gkougkousis, Evangelos Leonidas and Grieco, Chiara and Grinstein, Sebastian",
    title = "{Characterization of Irradiated Boron, Carbon-Enriched and Gallium Si-on-Si Wafer Low Gain Avalanche Detectors}",
    doi = "10.3390/instruments6010002",
    journal = "Instruments",
    volume = "6",
    number = "1",
    pages = "2",
    year = "2021"
}

@article{Ferrero:2018fen,
    author = "Ferrero, M. and others",
    title = "{Radiation resistant LGAD design}",
    eprint = "1802.01745",
    archivePrefix = "arXiv",
    primaryClass = "physics.ins-det",
    doi = "10.1016/j.nima.2018.11.121",
    journal = "Nucl. Instrum. Meth. A",
    volume = "919",
    pages = "16--26",
    year = "2019"
}

@article{Ali:2023roa,
    author = "Ali, S. and others",
    title = "{Performance in beam tests of carbon-enriched irradiated Low Gain Avalanche Detectors for the ATLAS High Granularity Timing Detector}",
    eprint = "2303.07728",
    archivePrefix = "arXiv",
    primaryClass = "physics.ins-det",
    doi = "10.1088/1748-0221/18/05/P05005",
    journal = "JINST",
    volume = "18",
    number = "05",
    pages = "P05005",
    year = "2023"
}

@article{Beresford:2023egc,
    author = "Beresford, L. A. and others",
    title = "{Destructive breakdown studies of irradiated LGADs at beam tests for the ATLAS HGTD}",
    eprint = "2306.12269",
    archivePrefix = "arXiv",
    primaryClass = "physics.ins-det",
    doi = "10.1088/1748-0221/18/07/P07030",
    journal = "JINST",
    volume = "18",
    number = "07",
    pages = "P07030",
    year = "2023"
}

@article{Jansen:2016bkd,
    author = "Jansen, Hendrik and others",
    title = "{Performance of the EUDET-type beam telescopes}",
    eprint = "1603.09669",
    archivePrefix = "arXiv",
    primaryClass = "physics.ins-det",
    reportNumber = "DESY-16-055",
    doi = "10.1140/epjti/s40485-016-0033-2",
    journal = "EPJ Tech. Instrum.",
    volume = "3",
    number = "1",
    pages = "7",
    year = "2016"
}

@article{Benoit_2016,
   title={The FE-I4 telescope for particle tracking in testbeam experiments},
   volume={11},
   ISSN={1748-0221},
   url={http://dx.doi.org/10.1088/1748-0221/11/07/P07003},
   DOI={10.1088/1748-0221/11/07/p07003},
   number={07},
   journal={Journal of Instrumentation},
   publisher={IOP Publishing},
   author={Benoit, M. and Mendizabal, J. Bilbao De and Bello, F.A. Di and Ferrere, D. and Golling, T. and Gonzalez-Sevilla, S. and Iacobucci, G. and Kocian, M. and Muenstermann, D. and Ristic, B. and Sciuccati, A.},
   year={2016},
   month=jul, pages={P07003–P07003} 
}

@article{Diener:2018qap,
    author = "Diener, R. and others",
    title = "{The DESY II Test Beam Facility}",
    eprint = "1807.09328",
    archivePrefix = "arXiv",
    primaryClass = "physics.ins-det",
    reportNumber = "DESY 18-111, DESY-18-111",
    doi = "10.1016/j.nima.2018.11.133",
    journal = "Nucl. Instrum. Meth. A",
    volume = "922",
    pages = "265--286",
    year = "2019"
}

@techreport{SPS_H6,
      author        = "Banerjee, Dipanwita and Bernhard, Johannes and Brugger,
                       Markus and Charitonidis, Nikolaos and Doble, Niels and
                       Gatignon, Lau and Gerbershagen, Alexander",
      title         = "{The North Experimental Area at the Cern Super Proton
                       Synchrotron}",
  institution   = {CERN},
  reportNumber  = {CERN-ACC-NOTE-2021-0015},
  year          = {2021},
  url           = {https://cds.cern.ch/record/2774716},
}

@article{Dannheim_2021,
   title={Corryvreckan: a modular 4D track reconstruction and analysis software for test beam data},
   volume={16},
   ISSN={1748-0221},
   url={http://dx.doi.org/10.1088/1748-0221/16/03/P03008},
   DOI={10.1088/1748-0221/16/03/p03008},
   number={03},
   journal={Journal of Instrumentation},
   publisher={IOP Publishing},
   author={Dannheim, D. and Dort, K. and Huth, L. and Hynds, D. and Kremastiotis, I. and Kröger, J. and Munker, M. and Pitters, F. and Schütze, P. and Spannagel, S. and Vanat, T. and Williams, M.},
   year={2021},
   month=mar, pages={P03008}
}

@article{Liu:2686241,
      author        = "Liu, Y. and Amjad, M.S. and Baesso, P. and Cussans, D. and
                       Dreyling-Eschweiler, J. and Ete, R. and Gregor, I. and
                       Huth, L. and Irles, A. and Jansen, H. and Krueger, K. and
                       Kvasnicka, J. and Peschke, R. and Rossi, E. and Rummler, A.
                       and Sefkow, F. and Stanitzki, M. and Wing, M. and Wu, M.",
      title         = "{EUDAQ2 -- A Flexible Data Acquisition Software Framework
                       for Common Test Beams}",
      archivePrefix = "arXiv",
      eprint        = "1907.10600",
      journal       = "JINST",
      volume        = "14",
      number        = "10",
      pages         = "P10033-P10033",
      year          = "2019",
      url           = "https://cds.cern.ch/record/2686241",
      note          = "17 pages, 6 figures",
      doi           = "10.1088/1748-0221/14/10/P10033",
}

@article{Galloway:2017gfx,
    author = "Galloway, Z. and others",
    title = "{Properties of HPK UFSD after neutron irradiation up to 6e15 n/cm$^2$}",
    eprint = "1707.04961",
    archivePrefix = "arXiv",
    primaryClass = "physics.ins-det",
    doi = "10.1016/j.nima.2019.05.017",
    journal = "Nucl. Instrum. Meth. A",
    volume = "940",
    pages = "19--29",
    year = "2019"
}

@article{Baesso:2019smg,
    author = "Baesso, P. and Cussans, D. and Goldstein, J.",
    title = "{The AIDA-2020 TLU: a flexible trigger logic unit for test beam facilities}",
    eprint = "2005.00310",
    archivePrefix = "arXiv",
    primaryClass = "physics.ins-det",
    doi = "10.1088/1748-0221/14/09/P09019",
    journal = "JINST",
    volume = "14",
    number = "09",
    pages = "P09019",
    year = "2019"
}

@article{Li:2022ukj,
    author = "Li, C. H. and Yang, X. and Ge, J. J. and Wang, T. and Zheng, X. X. and Sun, Y. J. and Liu, Y. W.",
    title = "{Performance of LGAD sensors with carbon enriched gain layer produced by USTC}",
    doi = "10.1016/j.nima.2022.167008",
    journal = "Nucl. Instrum. Meth. A",
    volume = "1039",
    pages = "167008",
    year = "2022"
}

@article{Li:2021ouc,
    author = "Li, Mengzhao and others",
    title = "{Effects of shallow carbon and deep N++ layer on the radiation hardness of IHEP-IME LGAD sensors}",
    eprint = "2110.12632",
    archivePrefix = "arXiv",
    primaryClass = "physics.ins-det",
    month = "10",
    year = "2021"
}

@article{Padilla:2020sau,
    author = "Padilla, R. and others",
    title = "{Effect of deep gain layer and Carbon infusion on LGAD radiation hardness}",
    eprint = "2004.05260",
    archivePrefix = "arXiv",
    primaryClass = "physics.ins-det",
    doi = "10.1088/1748-0221/15/10/P10003",
    journal = "JINST",
    volume = "15",
    number = "10",
    pages = "P10003",
    year = "2020"
}

@article{Snoj:2012dib,
    author = "Snoj, Luka and {\v{Z}}erovnik, Ga{\v{s}}per and Trkov, Andrej",
    title = "{Computational analysis of irradiation facilities at the JSI TRIGA reactor}",
    doi = "10.1016/j.apradiso.2011.11.042",
    journal = "Appl. Radiat. Isot.",
    volume = "70",
    pages = "483--488",
    year = "2012"
}

@article{Ambrozic:2017,
	author = "Ambro{\v{z}}i{\v{c}}, Klemen and {\v{Z}}erovnik, Ga{\v{s}}per and Snoj, Luka",    
    title = "{Computational analysis of the dose rates at JSI TRIGA reactor irradiation facilities}",
    doi = "10.1016/j.apradiso.2017.09.022",
    journal = "Appl. Radiat. Isot.",
    volume = "130",
    pages = "140--152",
    year = "2017"
}

@misc{Hamamatsu:R3809U-50,
  author       = {Hamamatsu Photonics K.K.},
  title        = {R3809U-50 Microchannel Plate Photomultiplier Tube},
  howpublished = {\url{https://www.hamamatsu.com/jp/en/product/optical-sensors/pmt/pmt_tube-alone/mcp-pmt/R3809U-50.html}}

}

@article{Jimenez-Ramos:2022,
  author = "Jiménez-Ramos, M. C. and others",
  title  = "Study of Ionization Charge Density-Induced Gain Suppression in LGADs",
  journal= "Sensors",
  year   = "2022",
  volume = "22",
  number = "3",
  pages  = "1080",
  doi    = "10.3390/s22031080"
}

@misc{IME:CAS_overview,
  author       = "{Institute of Microelectronics, Chinese Academy of Sciences (IME, CAS)}",
  title        = "{Institute Overview  IME, CAS}",
  howpublished = "\url{https://english.ime.cas.cn/au/Overview/}",
}

@article{article_electronDrift,
author = {Smith, P. and Inoue, M. and Frey, Jeffrey},
year = {1980},
month = {12},
pages = {797 - 798},
title = {Electron velocity in Si and GaAs at very high electric fields},
volume = {37},
journal = {Applied Physics Letters},
doi = {10.1063/1.92078}
}
